\documentclass[lettersize,journal]{IEEEtran}
\usepackage{amsmath,amsfonts}
\usepackage{algorithmic}
\usepackage{algorithm}
\usepackage{array}
\usepackage[caption=false,font=scriptsize,labelfont=sf]{subfig}
\usepackage{textcomp}
\usepackage{stfloats}
\usepackage{url}
\usepackage{verbatim}
\usepackage{graphicx}
\usepackage{cite}
\usepackage{bm}
\usepackage{accents}
\usepackage{array}
\usepackage{multicol}
\usepackage{multirow}
\usepackage{eqparbox}
\usepackage{enumerate}
\usepackage{xcolor}
\usepackage{makecell}

\newcolumntype{M}[1]{>{\centering\arraybackslash}m{#1}}
\newcolumntype{P}[1]{>{\centering\arraybackslash}p{#1}}

\newcommand\munderbar[1]{%
  \underaccent{\bar}{#1}}

\begin{document}

\title{Deep Reinforcement Learning for Voltage Control and Renewable Accommodation Using Spatial-Temporal Graph Information}

\author{Jinhao Li,
Ruichang Zhang,
Hao Wang, ~\IEEEmembership{Member~IEEE,}
~\\
Zhi Liu, ~\IEEEmembership{Member~IEEE,}
Hongyang Lai,
Yanru Zhang, ~\IEEEmembership{Member~IEEE}
\thanks{This work was supported by the National Natural Science Foundation of China (NSFC) under Grant 62001085, the General Program of Shenzhen Science and Technology Research and Development Fund under Grant JCYJ20220530164813029, and the Australian Research Council (ARC) Discovery Early Career Researcher Award (DECRA) under Grant DE230100046.}
\thanks{J. Li is with the School of Mechanical and Electrical Engineering, University of Electronic Science and Technology of China (UESTC), Chengdu, China, 611731. He is also with the Department of Data Science and AI, Faculty of Information Technology, Monash University, Melbourne, VIC 3800, Australia (e-mail: stephlee175@gmail.com)}
\thanks{R. Zhang is with the Department of Computer Science, University of Manchester (e-mail: zhangruichang2@gmail.com)}
\thanks{H. Wang is with the Department of Data Science and AI, Faculty of Information Technology, Monash University, Melbourne, VIC 3800, Australia. He is also affiliated with the Monash Energy Institute (e-mail: hao.wang2@monash.edu)}
\thanks{Z. Liu is with the Department of Computer and Network Engineering, University of Electro-Communications, Tokyo, Japan (e-mail:liu@ieee.org).}
\thanks{H. Lai is with the School of Mechanical and Electrical Engineering, University of Electronic Science and Technology of China, Chengdu, China, 611731 (e-mail: laihongyang0314@outlook.com).}
\thanks{Y. Zhang is with the University of Electronic Science and Technology of China (UESTC), Chengdu, China, 611731. She is also affiliated with Shenzhen Institute for Advanced Study of UESTC (e-mail:yanruzhang@uestc.edu.cn). Y. Zhang is the corresponding author.}
}

\markboth{Journal of \LaTeX\ Class Files,~Vol.~14, No.~8, August~2021}%
{Shell \MakeLowercase{\textit{et al.}}: A Sample Article Using IEEEtran.cls for IEEE Journals}


\maketitle
\begin{abstract}
Renewable energy resources (RERs) have been increasingly integrated into distribution networks (DNs) for decarbonization. However, the variable nature of RERs introduces uncertainties to DNs, frequently resulting in voltage fluctuations that threaten system security and hamper the further adoption of RERs. To incentivize more RER penetration, we propose a deep reinforcement learning (DRL)-based strategy to dynamically balance the trade-off between voltage fluctuation control and renewable accommodation. To further extract multi-time-scale spatial-temporal (ST) graphical information of a DN, our strategy draws on a multi-grained attention-based spatial-temporal graph convolution network (MG-ASTGCN), consisting of ST attention mechanism and ST convolution to explore the node correlations in the spatial and temporal views. The continuous decision-making process of balancing such a trade-off can be modeled as a Markov decision process optimized by the deep deterministic policy gradient (DDPG) algorithm with the help of the derived ST information. We validate our strategy on the modified IEEE 33, 69, and 118-bus radial distribution systems, with outcomes significantly outperforming the optimization-based benchmarks. Simulations also reveal that our developed MG-ASTGCN can to a great extent accelerate the convergence speed of DDPG and improve its performance in stabilizing node voltage in an RER-rich DN. Moreover, our method improves the DN's robustness in the presence of generator failures.
\end{abstract}


\section{Introduction} \label{sec:intro}
\IEEEPARstart{T}{here} has been an exponential growth of distributed renewable energy resources (RERs), e.g, wind and solar energy, in distribution networks (DNs) for mitigating global climate change and providing affordable electricity to customers~\cite{meinshausen2022}. From 2007 to 2021, the global installed capacity of solar photovoltaic (PV) has increased from 8 GW to 940 GW, so as the wind power (from 94 GW to 837 GW)~\cite{RER2021}. Despite various benefits brought by RERs, e.g., decarbonization and power supply cost reduction, the increasing adoption of RERs introduces considerable uncertainties to DNs due to their intermittent nature~\cite{IPCC2022}. Uncontrollable factors of RERs, such as solar irradiation and wind speed, can produce stochastic and non-dispatchable RER generation, which makes generation forecasts extremely challenging, leads to frequent voltage fluctuations, threatens system security, and may cause potentially economic losses~\cite{sinsel2020}.

Voltage control in an RER-rich DN has been widely discussed in the literature and can be basically sorted into three classes~\cite{plytaria2017}. 1) \textit{Distributed-optimization-based methods}: the optimal voltage control strategy is often derived from a non-convex optimal power flow problem and then relaxed into a centralized convex problem through semi-definite programming or second-order cone programming, which is solved by distributed algorithms. Typical consensus algorithms~\cite{bolognani2014,zhang2015,maknouninejad2014,utkarsh2016} and alternating direction method of multipliers~\cite{dall2013,zheng2016,robbins2016,robbins2016-2} are among the most prevalent solutions for distributed optimization problems. However, both methods suffer from heavy computational costs, in particular in large-scale DNs. 2) \textit{Decentralized-based methods}: based on network partitions, an optimization agent supported by conventional numerical algorithms~\cite{yu2012,di2013,dall2016} and heuristic methods (e.g., genetic algorithm~\cite{abessi2016}, particle swarm optimization~\cite{nayeripour2016}, harris hawk optimization~\cite{mahmoud2020_hho}, and grey wolf optimization~\cite{routray2020_gwo}) can be applied to each divided zone to achieve the overall voltage control in a decentralized manner. Nevertheless, numerical algorithms are challenging to converge in face of high uptake of RERs, while heuristic algorithms highly rely on accurate prior knowledge of a specific DN and can easily trap into local optimums. 3) \textit{Learning-based methods}: Unlike the aforementioned optimization-based methods, deep reinforcement learning (DRL)-based methods have drawn increasing attention to voltage control due to its model-free characteristic~\cite{wang2020,cao2020,yang2020,zhang2021,cao2021,sun2021,liu2021}, enabling the strategy to learn totally from historical experiences without any prior knowledge of both the uncertainty of renewables and the DN. Moreover, the DRL, benefiting from its interactive learning manner, is also well-suitable to capture the uncertain dynamics in an RER-rich DN and thus better control voltage fluctuations. However, learning a stable and well-performed DRL-based control strategy in complex physical systems is notoriously difficult due to its slow learning convergence, such as the voltage control problem in DNs. Furthermore, ensuring the increasing accommodation of renewables in DNs while mitigating voltage fluctuations has been inadequately discussed in the literature.

To bridge the research gap, we are motivated to propose a DRL-based strategy to balance the trade-off between voltage fluctuation control and renewable accommodation, leveraging spatial-temporal (ST) graphical information of the DN. Specifically, providing that correlations among node pairs in DN are mutually influenced in the spatial view while each node features are inherently dependent in the temporal view, we develop a novel multi-grained attention-based spatial-temporal convolution network (MG-ASTGCN) to explore ST correlations through ST attention mechanism and extract ST features through ST convolution in multiple time scales. The derived ST information indicates time-varying patterns of the DN's power flow, which can be potentially utilized by the DRL to accelerate its learning process, simultaneously improve its performance, and effectively interpret the graphical correlations among nodes in the DN. The main contributions of our work are summarized as follows.
\begin{itemize}
    \item \emph{ST Information Extraction}: We develop a novel MG-ASTGCN to fully extract ST information from an RER-rich DN graph. Specifically, attention mechanism and graph convolution inside the MG-ASTGCN are facilitated for exploring ST correlations and features, respectively. Moreover, since the power flow exhibits periodic patterns in multiple time scales, we construct multi-grained power flow time series to better capture temporal ST information.
    \item \emph{DRL for Balancing Voltage Control and Renewable Accommodation}: We propose a DRL-based strategy leveraging the derived ST graphical information to dynamically balance the trade-off between voltage fluctuation control and renewable accommodation. The consecutive control process is modeled as a Markov decision process (MDP) which is optimized via the cutting-edge off-policy DRL algorithm, namely the deep deterministic policy gradient (DDPG).
    \item \emph{Numerical Simulations and Implications}: We validate our DRL-based method on modified IEEE 33, 69, and 118-bus radial distribution systems. Simulations demonstrate the effectiveness of our approach with outcomes significantly outperforming benchmark algorithms, i.e., harris hawk optimization (HHO), grey wolf optimization (GWO), interior-point (IP)-based method, and linear/quadratic programming (LQP)-based method. Moreover, our DRL-based strategy improves network stability in the presence of generator failures, especially for large-scale DNs.
\end{itemize}

The key insights drawn from simulation results are summarized as follows.
\begin{itemize}
    \item \textit{DDPG converges faster with the assistance of MG-ASTGCN}: The developed MG-ASTGCN substantially accelerates the convergence speed of the DDPG, compared to other graphical correlation extraction methods, which demonstrates the effectiveness of our MG-ASTGCN in capturing the underlying ST graphical information of the DN.
    \item \textit{Effectiveness of the ST attention mechanism inside the MG-ASTGCN}: The spatial attention mechanism in the MG-ASTGCN captures mutual correlations among node pairs in the DN, while temporal attention exploits temporal correlations of each node features among multiple time steps. Simulations suggest that node pairs with more generator integration tend to have stronger spatial correlations, while each node features are highly self-correlated in historical recent time steps from the temporal view.
    \item \textit{Overemphasizing voltage fluctuation control or renewable accommodation undermines DDPG's performance}: Our DRL-based strategy encodes all objectives and operational constraints of the derived optimization problem into reward functions as feedback from the DN. Modeling results present that, if the reward function for voltage fluctuation control or renewable accommodation is overemphasized, the DDPG's performance correspondingly degenerates, resulting in sub-optimal operations of the DN, highlighting that striking an effective balance between these objectives is essential for the optimal control of the DN.
\end{itemize}

The remainder of this paper is organized as follows. Section \ref{sec:system_model} formulates an optimization problem balancing the trade-off between voltage fluctuation control and renewable accommodation, followed by Section \ref{sec:method}, where our DRL-based strategy, consisting of the MG-ASTGCN and DDPG, is proposed. Experimental results are presented in Section \ref{sec:exps}. Section \ref{sec:conclusion} concludes.

\begin{figure}[!t]
    \centering
    \includegraphics[width=\linewidth]{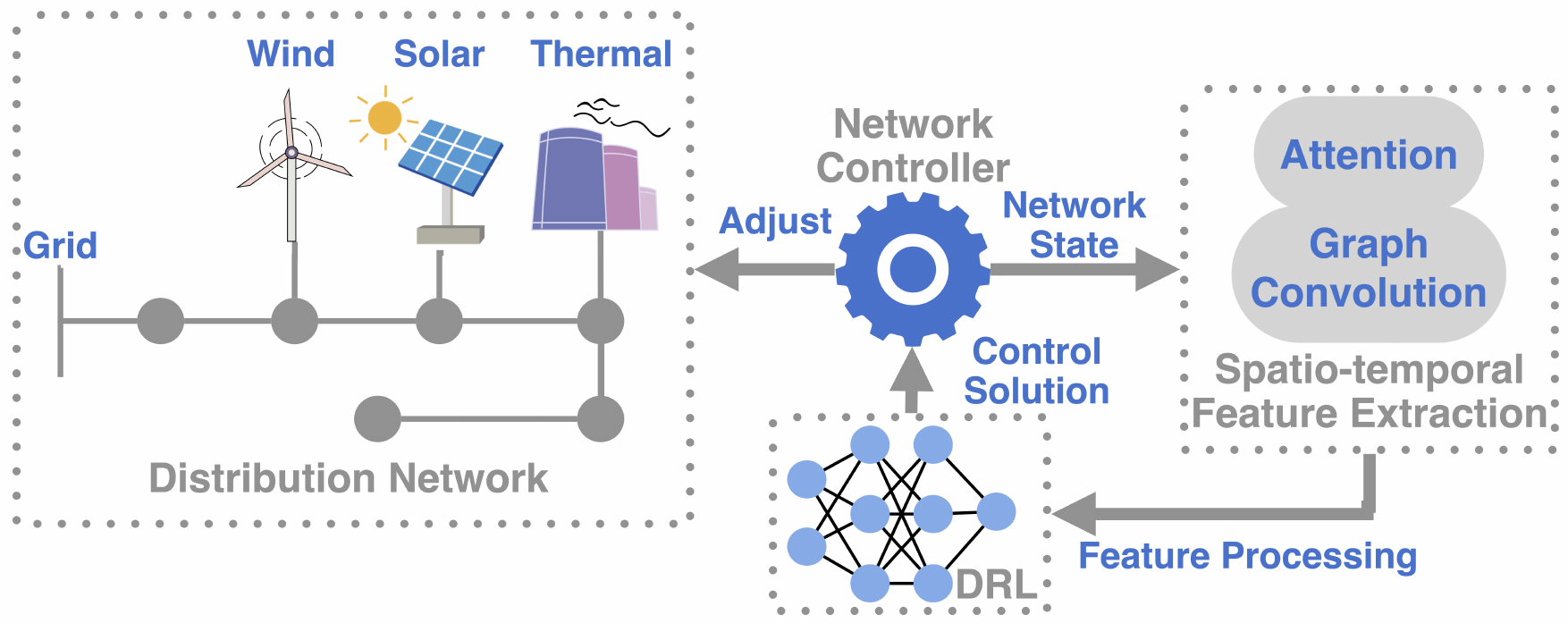}
    \caption{The system model and the presented work.}
    \label{fig:framework}
\end{figure}

\section{System Model} \label{sec:system_model}
We consider a radial DN highly integrated with RERs. We here utilize wind and solar PV generators as RERs since they are the most representative RERs with the largest installed capacities worldwide~\cite{RER2021}. The high uptake of RERs introduces considerable uncertainties in the DN and continuously leads to voltage fluctuations, hindering RER's further adoption and often causing curtailments. We formulate an optimization problem for mitigating voltage fluctuations while accommodating renewable integration, along with minimizing generation costs, from Section \ref{subsec:system_model_voltage_fluctuation} to \ref{subsec:system_model_renewable_accommdation}. An overview of the presented work is illustrated in Fig. \ref{fig:framework}.

\subsection{Voltage Fluctuation Control} \label{subsec:system_model_voltage_fluctuation}
The increasing presence of RERs frequently triggers voltage fluctuations, even voltage violations, at load nodes in the DN, which may severely degrade the performance of electronic equipment and pose potential security risks to electricity consumers. To mitigate voltage fluctuations, we consider voltage fluctuation control with an L2-norm-based metric $J^\text{vol}_t$ to describe voltage stability, which can be defined as
\begin{equation}
    \label{eq:voltage_flcutuation_editor_comment2}
    J^\text{vol}_t = \left[\sum_{n=1}^{N^\text{L}} \left( 1-\left|\mathbb{V}_{t,n}^\text{L}\right| \right)^2\right]^{\frac{1}{2}},
\end{equation}
where $t$ represents the current time step, $N^\text{L}$ is the number of load nodes, and $\mathbb{V}_n$ is the phasor form of load voltage expressed as $\mathbb{V}_{t,n}^\text{L} = v_{t,n}^\text{L}\angle \delta_{t,n}^\text{L}$, where $v_{t,n}^\text{L}$ is the voltage magnitude and $\delta_{t,n}^\text{L}$ is the phase angle.

\subsection{Renewable Accommodation} \label{subsec:system_model_renewable_accommdation}
Accommodating the increasing penetration of RERs is of great importance for an orderly energy transition in the power grid and functions as the main pillar for net-zero emissions. The metric reflecting the effectiveness of renewable accommodation, denoted by $J_t^\text{RER}$, can be formulated as
\begin{equation}
    \label{eq:renewable_accommdation}
    J_t^\text{RER} = \sum_{j=1}^{N^\text{W}} \frac{p_{t,j}^\text{W,act}}{\bar{p}_{t,j}^\text{W}} + \sum_{k=1}^{N^\text{S}}\frac{p_{t,k}^\text{S,act}}{\bar{p}_{t,k}^\text{S}},
\end{equation}
where $N^\text{W}$ and $N^\text{S}$ are the number of wind and solar PV generators, $p_{t,j}^\text{W,act}$ and $p_{t,k}^\text{S,act}$ are the actual outputs of wind and solar PV generation while $\bar{p}_{t,j}^\text{W}$ and $\bar{p}_{t,k}^\text{S}$ are correspondingly maximum power outputs at the current time step, which are usually obtained via onsite monitoring devices.

Moreover, the inherent variability of RERs can lead to mismatch costs between scheduled power generation and actual power output~\cite{panda2015}, which can be further divided into reserve cost (under power overestimation circumstance) and penalty cost (under power underestimation circumstance). The reserve costs for wind and solar PV power can be formulated as
\begin{equation}
\label{eq:wind_reserve_cost}
    \begin{aligned}
        C^\text{W,r}_{t,j} &=\mathbb{E}\left[c^{\text{W,r}}_{j}\left(\bar{p}_{t,j}^\text{W}-p^\text{W,ava}_{t,j}\right)\right],\\
        &=c^{\text{W,r}}_{j} \int_{\munderbar{p}_{t,j}^\text{W}}^{\bar{p}_{t,j}^\text{W}} \left(\bar{p}_{t,j}^\text{W}-p_{t,j}^\text{W,ava}\right) f^\text{W}\left(p_{t,j}^\text{W,ava}\right)d\left(p_{t,j}^\text{W,ava}\right),\\
    \end{aligned}   
\end{equation}
\begin{equation}
\label{eq:solar_reserve_cost}
    \begin{aligned}
    C^{\text{S,r}}_{t,k} &= \mathbb{E}\left[c^{\text{S,r}}_{k} \left(\bar{p}_{t,k}^\text{S}-p^{\text{S,ava}}_{t,k}\right)\right],\\
    &=c^{\text{S,r}}_{k} \int_{\munderbar{p}_{t,k}^\text{S}}^{\bar{p}_{t,k}^\text{S}} \left(\bar{p}_{t,k}^\text{S}-p_{t,k}^\text{S,ava}\right)f^\text{S}\left(p_{t,k}^\text{S,ava}\right) d\left(p_{t,k}^\text{S,ava}\right), 
    \end{aligned}
\end{equation}
where $j$ and $k$ are indices of wind and solar PV generators, $c^{\text{W,r}}_{j}$ and $c^{\text{S,r}}_{k}$ are constant coefficients, $p^\text{W,ava}_{t,j}$ and $p^{\text{S,ava}}_{t,k}$ are the available power outputs of wind and solar PV generation, $\munderbar{p}_{t,j}^\text{W}$ and $\munderbar{p}_{t,k}^\text{S}$ are the minimum power outputs of corresponding generators, $f^\text{W}(\cdot)$ and $f^\text{S}(\cdot)$ are the probability density functions of wind and solar PV generation, where we assume the wind speed and solar irradiation conform with the Weibull and lognormal distribution, respectively~\cite{panda2015}.

Similarly, the penalty costs of wind and solar PV generators can be formulated as
\begin{align}
\label{eq_wind-penalty-cost}
C^\text{W,p}_{t,j} &= \mathbb{E}\left[c^\text{W,p}_{j}\left(p^\text{W,ava}_{t,j}-\bar{p}_{t,j}^\text{W}\right)\right],\\
\label{eq_solar-penalty-cost}
C^\text{S,p}_{t,k} &= \mathbb{E}\left[c^\text{S,p}_{k}\left(p^\text{S,ava}_{t,k}-\bar{p}_{t,k}^\text{S}\right)\right],
\end{align}
where $c^\text{W,p}_{j}$ and $c^\text{S,p}_{k}$ are constant coefficients.

Additionally, fuel energy resources still play an irreplaceable role in maintaining DN's stability, especially in face of RER generation shortage. The generation costs of thermoelectric generators can be formulated as
\begin{equation}
\label{eq:thermal_gen_cost}
    C_{t,i}^\text{T} = a_i\left(p_{t,i}^\text{T}\right)^2+b_ip_{t,i}^\text{T}+c_i,
\end{equation}
where $i$ is the index of the thermoelectric generator, $a_i$, $b_i$, and $c_i$ are constant coefficients, $p_{t,i}^\text{T}$ is the actual power output of the thermoelectric generator.

Combining the generation costs of thermoelectric and RER generators, the overall generation cost in the DN at one time step can be formulated as
\begin{equation}
    \label{eq:gen_cost}
    J_t^\text{gen}=\sum_{i=1}^{N^\text{T}}C_{t,i}^\text{T} +\sum_{j=1}^{N^\text{W}} \left(C^\text{W,r}_{t,j}+C^\text{W,p}_{t,j}\right)+\sum_{k=1}^{N^\text{S}}\left(C^\text{S,r}_{t,k}+C^\text{S,p}_{t,k}\right),
\end{equation}
where $N^\text{T}$ is the number of thermoelectric generators.

\subsection{Optimization Formulation} \label{subsec:system_model_optimization_formulation}
We formulate the objectives of our control strategy as the weighted summation of metrics regarding voltage fluctuation mitigation, renewable accommodation maximization, and generation cost minimization, which can be expressed as
\begin{equation}
    \label{eq:optimization_obj}
    \begin{aligned}
    \min \hspace{0.25em}\sum_{t=1}^T &\left(\frac{w^\text{vol}}{N^\text{L}}J_t^\text{vol} - \frac{w^\text{RER}}{N^\text{W}+N^\text{S}}J_t^\text{RER}\right.,\\
    &\left. + \frac{w^\text{gen}}{N^\text{T}+N^\text{W}+N^\text{S}}J_t^\text{gen}\right),
    \end{aligned}
\end{equation}
where $T$ is the time horizon, $w^\text{vol}$, $w^\text{RER}$, and $w^\text{gen}$ are weights for their corresponding objectives, i.e., $J_t^\text{vol}$, $J_t^\text{RER}$, and $J_t^\text{gen}$. The objective defined in Eq. \eqref{eq:optimization_obj} is subject to physical constraints that ensure safe operations of the DN. In particular, the equality constraints, namely power balance equations of active and reactive power in the DN, are formulated as
\begin{equation}
    \label{eq:cons_active_power_balance}
    \sum_{i=1}^{N^\text{T}} p_{t,i}^\text{T} + \sum_{j=1}^{N^\text{W}} p_{t,j}^\text{W,act} + \sum_{k=1}^{N^\text{S}} p_{t,k}^\text{K,act} + p_t^\text{G} = \sum_{n=1}^{N^\text{L}} p_{t,n}^\text{L} + L_t^P,
\end{equation}
\begin{equation}
    \label{eq:cons_reactive_power_balance}
    \sum_{i=1}^{N^\text{T}} q_{t,i}^\text{T} + \sum_{j=1}^{N^\text{W}} q_{t,j}^\text{W,act} + \sum_{k=1}^{N^\text{S}} q_{t,k}^\text{K,act} + q_t^\text{G} = \sum_{n=1}^{N^\text{L}} q_{t,n}^\text{L} + L_t^Q,
\end{equation}
with the active/reactive power losses $L_t^P/L_t^Q$ on the branch defined as
\begin{equation}
    \label{eq:active_power_loss}
    L_t^P= \sum_{m,n}^{N^\text{B}} g_{m,n}\left[v_{t,m}^2+v_{t,n}^2-2v_{t,m}v_{t,n}\cos\left(\delta_{t,mn}\right)\right],
\end{equation}
\begin{equation}
    \label{eq:reactive_power_loss}
    L_t^Q= \sum_{m,n}^{N^\text{B}} b_{m,n}\left[v_{t,m}^2+v_{t,n}^2-2v_{t,m}v_{t,n}\cos\left(\delta_{t,mn}\right)\right],
\end{equation}
where $g_{m,n}$ and $b_{m,n}$ are the conductance and susceptance of the branch, $p_t^G$ and $q_t^G$ are the active/reactive power from the power grid.

Moreover, inequality constraints that limit both load and generator voltages can be defined as
\begin{align}
\label{eq:cons_load_voltage}
\munderbar{v}_{n}^\text{L} &\leq v_{t,n}^\text{L} \leq \bar{v}_{n}^\text{L}, && n=1,\cdots,N^\text{L},\\
\label{eq:cons_gen_voltage_thermal}
\munderbar{v}_{i}^\text{T} &\leq v_{t,i}^\text{T} \leq \bar{v}_{i}^\text{T}, && i=1,\cdots,N^\text{T},\\
\label{eq:cons_gen_voltage_wind}
\munderbar{v}_{j}^\text{W} &\leq v_{t,j}^\text{W} \leq \bar{v}_{j}^\text{W}, && j=1,\cdots,N^\text{W},\\
\label{eq:cons_gen_voltage_solar}
\munderbar{v}_{k}^\text{S} &\leq v_{t,k}^\text{S} \leq \bar{v}_{k}^\text{S}, && k=1,\cdots,N^\text{S},
\end{align}
where $\munderbar{v}_{n}^\text{L}$, $\munderbar{v}_{i}^\text{T}$, $\munderbar{v}_{j}^\text{W}$, $\munderbar{v}_{k}^\text{S}$ and $\bar{v}_{n}^\text{L}$, $\bar{v}_{i}^\text{T}$, $\bar{v}_{j}^\text{W}$, $\bar{v}_{k}^\text{S}$ are the minimum and maximum voltages of load nodes and generators, respectively.

The active and reactive power output limits of generators can be defined as
\begin{align}
\label{eq:cons_gen_active_power_thermal}
\munderbar{p}_{i}^\text{T} &\leq p_{t,i}^\text{T} \leq \bar{p}_{i}^\text{T}, && i=1,\cdots,N^\text{T},\\
\label{eq:cons_ramp_rate_thermal}
-p_i^\text{T,Down} &\leq p_{t+1,i}^\text{T}-p_{t,i}^\text{T}\leq p_i^\text{T,Up} , && i=1,\cdots,N^\text{T},\\
\label{eq:cons_gen_active_power_wind}
\munderbar{p}_{j}^\text{W} &\leq p_{t,j}^\text{W} \leq \bar{p}_{j}^\text{W}, && j=1,\cdots,N^\text{W},\\
\label{eq:cons_gen_active_power_solar}
\munderbar{p}_{k}^\text{S} &\leq p_{t,k}^\text{S} \leq \bar{p}_{k}^\text{S}, && k=1,\cdots,N^\text{S},\\
\label{eq:cons_gen_reactive_power_thermal}
\munderbar{q}_{i}^\text{T} &\leq q_{t,i}^\text{T} \leq \bar{q}_{i}^\text{T}, && i=1,\cdots,N^\text{T},\\
\label{eq:cons_gen_reactive_power_wind}
\munderbar{q}_{j}^\text{W} &\leq q_{t,j}^\text{W} \leq \bar{q}_{j}^\text{W}, && j=1,\cdots,N^\text{W},\\
\label{eq:cons_gen_reactive_power_solar}
\munderbar{q}_{k}^\text{S} &\leq q_{t,k}^\text{S} \leq \bar{q}_{k}^\text{S}, && k=1,\cdots,N^\text{S},
\end{align}
where $\munderbar{p}_{i}^\text{T}$, $\munderbar{p}_{j}^\text{W}$, $\munderbar{p}_{k}^\text{S}$ and $\bar{p}_{i}^\text{T}$, $\bar{p}_{j}^\text{W}$, $\bar{p}_{k}^\text{S}$ are the minimum and maximum active power outputs while $\munderbar{q}_{i}^\text{T}$, $\munderbar{q}_{j}^\text{W}$, $\munderbar{q}_{k}^\text{S}$ and $\bar{q}_{i}^\text{T}$, $\bar{q}_{j}^\text{W}$, $\bar{q}_{k}^\text{S}$ are the minimum and maximum reactive power outputs. Constraint \eqref{eq:cons_ramp_rate_thermal} describes the active power ramp rate constraint of the thermoelectric generators, where $p_i^\text{T,Down}$ and $p_i^\text{T,Up}$ are the ramp-down and ramp-up limits of the $i$-th thermoelectric generator, which are set as $25\%$ of its rated power by default~\cite{ma2019,woo2020}. Moreover, for the shut-down constraint of the thermoelectric generator, before the shut-down, the thermoelectric generator must adjust its current power output to its lower limit $\munderbar{p}_i^\text{T}$. Once shut down, the thermoelectric generator is not allowed to restart within $4$ time steps by default. For the start-up constraint, the unavailable thermoelectric generator must adjust its power output to its lower limit before connecting to the DN again. Furthermore, once functioning in the DN, the generator is not allowed to be shut down within $4$ time steps by default.

Also, the power flow constraint on the branch can be expressed as
\begin{equation}
\label{eq:cons_branch_flow}
|\mathbb{S}_{t,b}^\text{B}| \leq \bar{s}_{b}^\text{B}, b=1,\cdots,N^\text{B},
\end{equation}
where $b$ is the index of the branch, $\mathbb{S}_{t,b}^\text{B}=p_{t,b}^\text{B}+jq_{t,b}^\text{B}$ is the complex form of apparent power on the branch, and $\bar{s}_{b}^\text{B}$ represents the maximum power flow on the branch.

\section{Methodology} \label{sec:method}
To solve the optimization problem derived from Eq. \eqref{eq:optimization_obj} to \eqref{eq:cons_branch_flow}, we first develop the MG-ASTGCN to extract ST graphical information of the DN in Section \ref{subsec:method_ASTGCN}, with the aim of delivering prior graphical information of the DN for its sequential DDPG. The consecutive control problem is modeled as an MDP in Section \ref{subsec:method_DRL}, where we introduce the DDPG to learn the optimal strategy for controlling voltage fluctuations and accommodating renewable generation.

\subsection{ST Information Extraction via MG-ASTGCN} \label{subsec:method_ASTGCN}
\subsubsection{MG-ASTGCN Preliminaries} \label{subsubsec:methd_ASTGCN_preliminary}
\begin{figure}[!t]
    \centering
    \includegraphics[width=\linewidth]{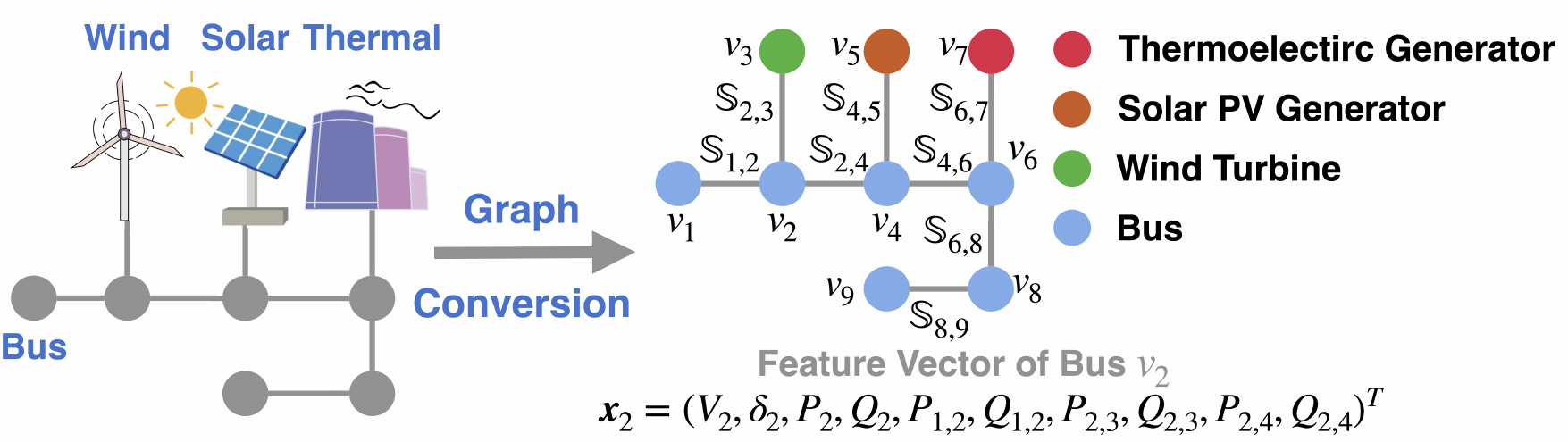}
    \caption{The undirected graph of a DN.}
    \label{fig:dn_to_graph}
\end{figure}

A DN can be modeled as an undirected graph $\mathcal{G}=(\mathcal{V},\mathcal{E},\bm{A})$, as illustrated in Fig. \ref{fig:dn_to_graph}, where $\mathcal{V}$, $\mathcal{E}$, and $\bm{A}$ represent the node set, edge set, and the adjacency matrix, respectively. Each load node $v_n$ generates a feature vector $\bm{x}_{t,n}$ for gathering local information, e.g., load voltage ($\mathbb{V}_{t,n}^\text{L}=v_{t,n}^\text{L}\angle\delta_{t,n}^\text{L}$), load power ($p_{t,n}^\text{L}, q_{t,n}^\text{L}$), and connected branch power ($\mathbb{S}_{t,b}^\text{B}=p_{t,b}^\text{B}+jq_{t,b}^\text{B}$), at each time step, which can be formulated as
\begin{equation}
\label{eq:node_feature_vec}
\hspace{-0.15em}\bm{x}_{t,n} = \left[v_{t,n}^\text{L}, \delta_{t,n}^\text{L}, p_{t,n}^\text{L}, q_{t,n}^\text{L}, p_{t,b1}^\text{B}, q_{t,b1}^\text{B},\cdots\right]^T\in \mathbb{R}^{F_n\times 1},
\end{equation}
where $F_n$ is the dimension of node features. The feature matrix of graph $\mathcal{G}$ can be aggregated as
\begin{equation}
    \label{eq:graph_feature_vec}
    \bm{X}_t = \left[\bm{x}_{t,1},\bm{x}_{t,2},\cdots,\bm{x}_{t,N^\text{L}}\right]\in \mathbb{R}^{F'\times N^\text{L}},
\end{equation}
where $F'$ is the highest node feature dimension defined as $F'=\max_{n}F_n$.

Considering periodic patterns in the power flow, especially the daily and weekly ones \cite{RER2021}, we develop a multi-grained vector constructor to better capture temporal correlations of the DN, where the recent, daily, and weekly graph segments can be formulated as
\begin{align}
\label{eq:recent_segment}
\bm{\mathcal{X}}^\text{r} &= \left[\bm{X}_{t-T^\text{r}+1},\cdots,\bm{X}_{t-1},\bm{X}_{t}\right]\in\mathbb{R}^{F'\times N^\text{L}\times T^\text{r}},\\
\label{eq:daily_segment}
\bm{\mathcal{X}}^\text{d} &= \left[\bm{X}_{t-T^\text{d}\times N^{\text{d}}},\cdots,\bm{X}_{t-N^{\text{d}}},\bm{X}_{t}\right]\in\mathbb{R}^{F'\times N^\text{L}\times T^\text{d}},\\
\label{eq:weekly_segment}
\bm{\mathcal{X}}^\text{w} &= \left[\bm{X}_{t-7\times T^\text{w}\times N^{\text{d}}},\cdots,\bm{X}_{t-7\times N^{\text{d}}},\bm{X}_{t}\right]\in\mathbb{R}^{F'\times N^\text{L}\times T^\text{w}},
\end{align}
where $T^\text{r}$, $T^\text{d}$, and $T^\text{w}$ indicate the length of recent, daily, and weekly segments, respectively, and $N^\text{d}$ represents the number of resolving the optimization problem per day.

\begin{figure}[!t]
    \centering
    \includegraphics[width=\linewidth]{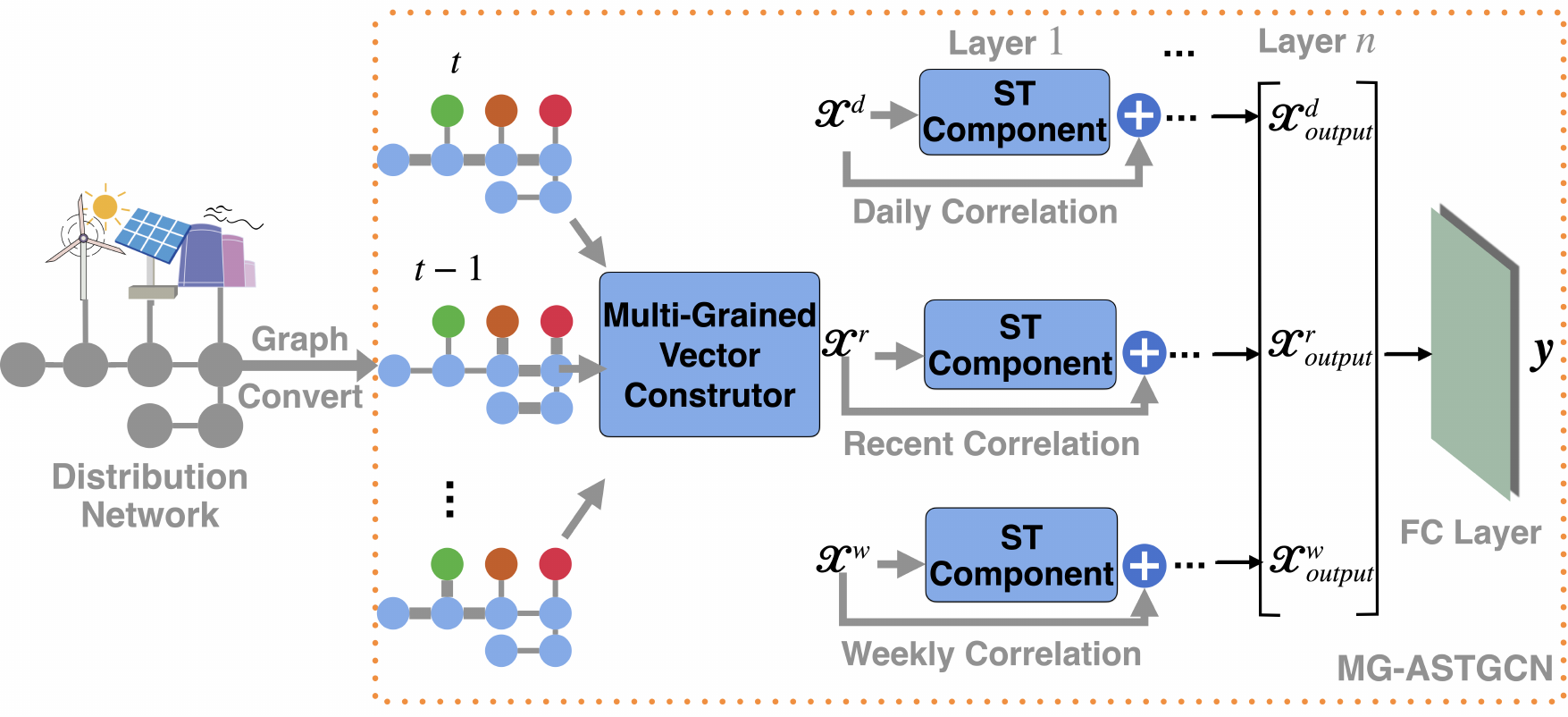}
    \caption{The framework of the proposed MG-ASTGCN.}
    \label{fig:framework_MGASTGCN}
\end{figure}

The framework of the proposed MG-ASTGCN is illustrated in Fig. \ref{fig:framework_MGASTGCN}, consisting of graph conversion operation, multi-grained vector constructor, and the most essential ASTGCN which takes multi-grained segments as inputs and employs stacked ST components to extract ST information. The structure of one ST component is depicted in Fig. \ref{fig:structure_ST_component}, including ST attention mechanism and ST convolution, which are presented in detail in Section \ref{subsubsec:method_ASTGCN_ST_attention} and \ref{subsubsec:method_ASTGCN_ST_conv}, respectively.

\subsubsection{Spatial-Temporal Attention Mechanism} \label{subsubsec:method_ASTGCN_ST_attention}
The key idea of the ST attention mechanism is to pay more attention to valuable graphical information in both spatial and temporal perspectives, assisting its sequential ST convolution for extracting more useful features for the DRL.

\begin{figure}[!t]
    \centering
    \includegraphics[width=\linewidth]{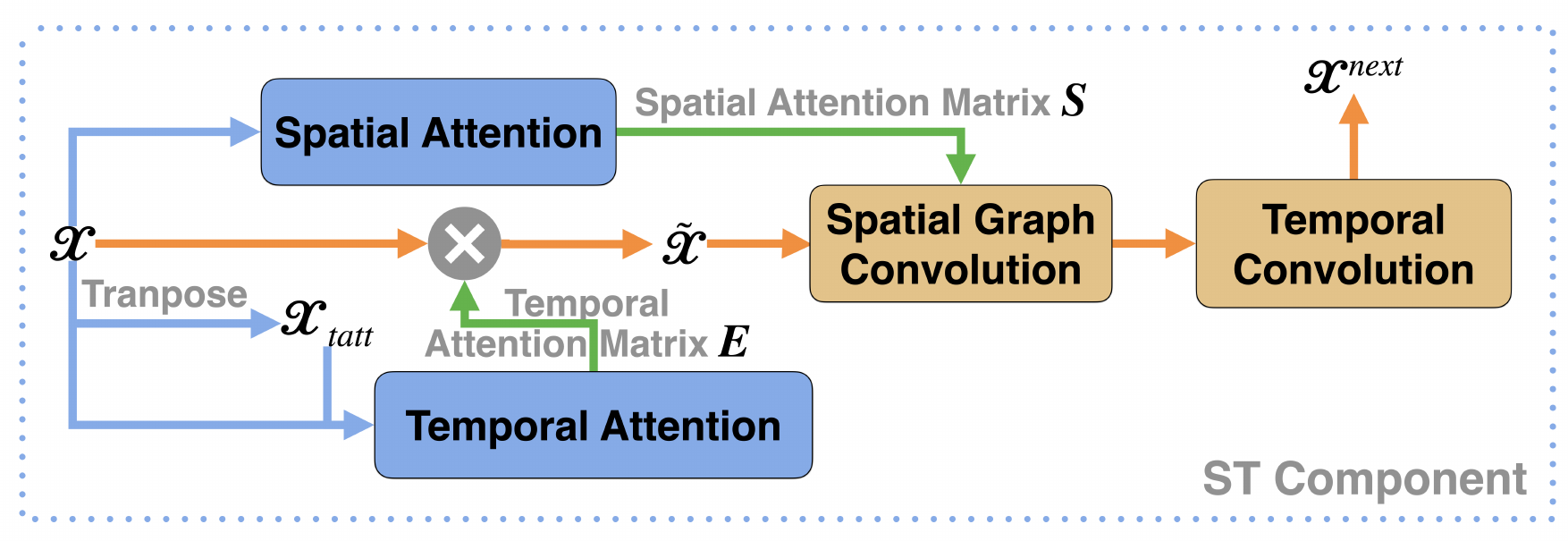}
    \caption{The structure of one ST component.}
    \label{fig:structure_ST_component}
\end{figure}

\textbf{Spatial Attention}: Mutual influences of each neighboring node pair in the DN dynamically vary due to the changes in power flow. To explore such mutual influences, an attention mechanism in spatial dimension is thus developed to capture the dynamic correlations~\cite{shi2021}, which can be formulated as (we omit the notations of multi-grained segments in the rest of this section for brevity)
\begin{gather}
\label{eq:spatial_att}
\bm{S} = \bm{V}_s\odot \sigma\left[\left(\bm{\mathcal{X}}  \bm{W}_T\right)^T\bm{W}_{FT}\left(\bm{W}_F\bm{\mathcal{X}}\right)^T+\bm{b}_s\right],\\
\label{eq:spatial_att_softmax}
\bm{S} \leftarrow \text{Softmax}\left(\bm{S}\right),
\end{gather}
where $\odot$ represents element-wise multiplication, $\sigma(\cdot)$ is the sigmoid activation function, $\bm{V}_s$, $\bm{W}_T$, $\bm{W}_{FT}$, $\bm{W}_F$, and $\bm{b}_s$ are all learnable parameters, and $\bm{S}$ represents the spatial attention matrix, whose element $s_{i,j}$, namely the attention weight, semantically describes the correlation strength between the $i$-th and $j$-th nodes. The derived spatial attention matrix can be
adopted for spatial graph convolution to adjust spatial correlation strengths of node pairs, as shown in Fig. \ref{fig:structure_ST_component}.

\textbf{Temporal Attention}: Similar to the spatial attention, the temporal attention mechanism \cite{vaswani2017} aims to track temporal correlations of the changing node features, which can be formulated as
\begin{gather}
\label{eq:temporal_att}
\bm{E} = \bm{V}_e\odot \sigma\left[\left(\bm{U}_N \bm{\mathcal{X}}_{\text{tatt}}\right)^T\bm{U}_{FN}\left(\bm{U}_F\bm{\mathcal{X}}\right)+\bm{b}_e\right],\\
\label{eq:temporal_att_softmax}
\bm{E} \leftarrow \text{Softmax}\left(\bm{E}\right),
\end{gather}
where $\bm{V}_e$, $\bm{U}_N$, $\bm{U}_{FN}$, $\bm{U}_F$, and $\bm{b}_e$ are learnable parameters, $\bm{\mathcal{X}}_{\text{tatt}}$ is the transposed form of the input segment $\bm{\mathcal{X}}$, and $\bm{E}$ is referred to as the temporal attention matrix whose attention weight $e_{i,j}$ represents the temporal dependencies between two graph feature vectors $\bm{X}_{t-i}$ and $\bm{X}_{t-j}$. The temporal attention matrix is used to add temporal correlation information to the original input segment $\bm{\mathcal{X}}$ as shown in Fig. \ref{fig:structure_ST_component}, which can be expressed as $\tilde{\bm{\mathcal{X}}} = \bm{\mathcal{X}}\bm{E}$.

\subsubsection{Spatial-Temporal Convolution} \label{subsubsec:method_ASTGCN_ST_conv}
The ST convolution consists of spatial graph convolution and temporal convolution, aiming to compress the three-dimensional input segments and extract ST features that can be integrated with the DRL algorithm.

\textbf{Spatial Graph Convolution}: Graph convolution is defined as a convolution operation implemented by using linear operators diagonalizing in the Fourier domain to replace the classical convolution operator \cite{zhang2019}, which can be expressed as
\begin{equation}
\label{eq:spatial_graph_conv}
    \begin{aligned}
    \text{ReLU}\left(g_\theta *_G \tilde{\bm{\mathcal{X}}}\right) &= \text{ReLU}\left[g_\theta \left(\bm{L}\right) \tilde{\bm{\mathcal{X}}}\right],\\
    &= \text{ReLU}\left[\bm{\Lambda}^T\left(\bm{\Lambda}\tilde{\bm{\mathcal{X}}}\odot\bm{\Lambda}g_\theta\right)\right],
    \end{aligned}
\end{equation}
where $*_G$ represents the graph convolution operator, $g_\theta$ is a convolution filter, $\bm{L}$ is the Laplacian matrix of the derived undirected graph $\mathcal{G}$, $\bm{\Lambda}$ is the result of eigenvalue decomposition, and the rectified linear unit (ReLU) is adopted as the activation function. Providing that conducting eigenvalue decomposition on the Laplacian matrix is computationally expensive, Chebyshev polynomials are often used for approximating the solution of eigenvalue decomposition in practice~\cite{simonovsky2017}. With the addition of spatial correlation information from the spatial attention matrix $\bm{S}$ as shown in Fig. \ref{fig:structure_ST_component}, the spatial graph convolution can be rewritten as
\begin{equation}
\label{eq:spatial_graph_conv_rewrite}
    \text{ReLU}\left(g_\theta *_G \tilde{\bm{\mathcal{X}}}\right) \approx \sum_{k=0}^{K-1} \theta_k\left[T_k(\tilde{\bm{L}})\odot \bm{S}\right]x,
\end{equation}
where $\tilde{\bm{L}}$ is the normalized Laplacian matrix, $\theta_k$ is the coefficient of Chebyshev polynomials, $K$ is the highest order of Chebyshev polynomials, and $T_k$ is the $k$-th order Chebyshev polynomial.

\textbf{Temporal Convolution and Feature Compression}:
The temporal convolution operation takes the result of the spatial graph convolution as input and performs convolution along the temporal dimension, i.e., $T^\text{r}$, $T^\text{d}$, and $T^\text{w}$, which can be formulated as
\begin{equation}
\label{eq:temporal_conv}
    \bm{\mathcal{X}}^\text{next} = \text{ReLU}\left\{h_\theta * \left[\text{ReLU}\left(g_\theta *_G \tilde{\bm{\mathcal{X}}}\right)\right]\right\},
\end{equation}
where $h_\theta$ is the temporal convolution filter and $\bm{\mathcal{X}}^\text{next}$ is the input for the following ST component.

To pass the extracted ST information to the DRL algorithm, the multi-time-scale outputs are fused and then fed into a fully-connected neural network layer (FCNNL)~\cite{goodfellow2016} for feature compression as shown in Fig. \ref{fig:framework_MGASTGCN}, which can be formulated as
\begin{equation}
\label{eq:output_ASTGCN}
    \bm{y} = \text{ReLU}\left\{\textbf{FCNNL}\left[\textbf{concat}\left(\bm{\mathcal{X}}^\text{r},\bm{\mathcal{X}}^\text{d},\bm{\mathcal{X}}^\text{w}\right)\right]\right\},
\end{equation}
where \textbf{concat} represents the concatenating operation for the multi-time-scale outputs.

\begin{figure*}[!t]
    \centering
    \includegraphics[width=0.75\linewidth]{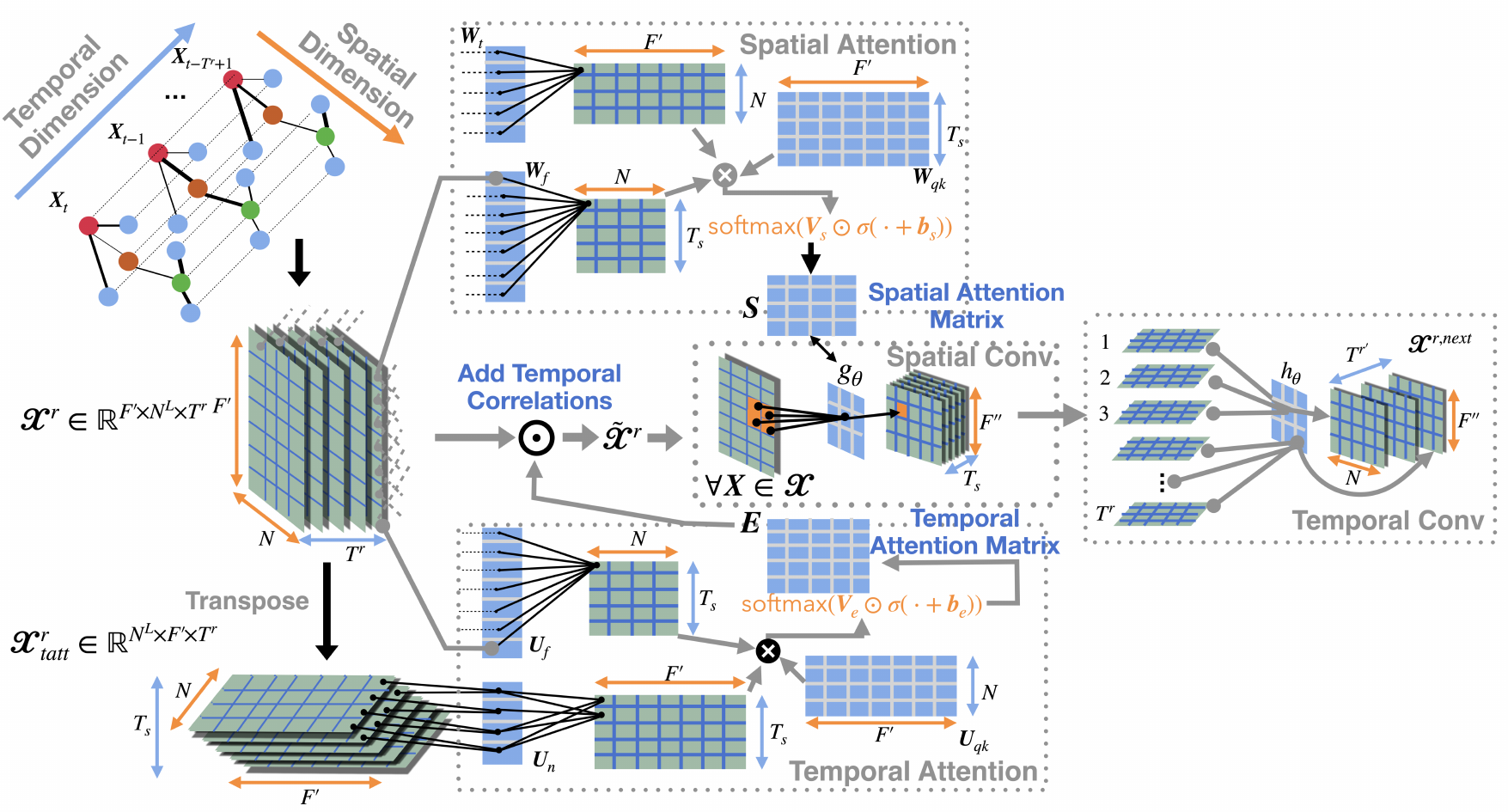}
    \caption{The detailed process of one recent graph segment passing through one whole ST component.}
    \label{fig:process_segment}
\end{figure*}

In summary, ST graphical information can be fully exploited by our MG-ASTGCN. The spatial attention mechanism explores spatial correlations of neighboring node pairs, while the temporal attention focuses on mining self-correlations of node features in the temporal view. Furthermore, based on ST correlations provided by the preceding ST attention, ST convolution extracts ST features of the underlying DN. The detailed process of the recent graph segment $\bm{\mathcal{X}}^\text{r}$ passing through one ST component is illustrated in Fig. \ref{fig:process_segment}.

\subsection{DRL-based Control Strategy} \label{subsec:method_DRL}
\subsubsection{MDP Modeling} \label{subsubec:method_DRL_MDP}
Balancing the trade-off between voltage control and renewable accommodation in the DN can be considered as a consecutive decision-making process, which can be further modeled as an MDP~\cite{sutton2018} consisting of four critical parts: state space $\mathcal{S}$, action space $\mathcal{A}$, probability space $\mathcal{P}$, and reward space $\mathcal{R}$.

\textbf{State Space $\mathcal{S}$}: The state of the $n$-th load node, denoted by $\bm{s}_{t,n}$, is its feature vector $\bm{x}_{t,n}$ defined in Eq. \eqref{eq:node_feature_vec}. Moreover, the extracted ST features $\bm{y}_t$ is aggregated with the DN's state. Thus, the state of the DN can be expressed as
\begin{equation}
    \label{eq:state}
    \bm{s}_t = \left[\bm{s}_{t,1},\cdots,\bm{s}_{t,N^\text{L}},\bm{y}_t\right].
\end{equation}

\textbf{Action Space $\mathcal{A}$}: For each power generator, only its active power $p_{t,i}$ and voltage magnitude $v_{t,i}$ can be manipulated. Thus, the actions of the $i$-th power generator can be expressed as $\bm{a}_{t,i}=\left[p_{t,i}, v_{t,i}\right]$. Actions of all generators in the DN can be defined as
\begin{equation}
    \label{eq:action}
    \bm{a}_t = \left[\bm{a}_{t,1}^\text{T},\cdots,\bm{a}_{t,N^\text{T}}^\text{T},\bm{a}_{t,1}^\text{W},\cdots,\bm{a}_{t,N^\text{W}}^\text{W},\bm{a}_{t,1}^\text{S},\cdots,\bm{a}_{t,N^\text{S}}^\text{S}\right],
\end{equation}
where $\bm{a}_{t,i}^\text{T}, \bm{a}_{t,j}^\text{W}, \bm{a}_{t,k}^\text{S}$ are actions of thermoelectric, wind, and solar PV generators. Note that, our proposed strategy is also applicable to generators operating in the P/Q control mode by changing the generator's action space into active power $p_{t,i}$ and reactive power $q_{t,i}$.

We would like to clarify that our proposed DRL-based strategy can be easily applicable to generators operating in the P/Q control mode by slightly modifying the generator's action space. In our original manuscript, the action space of each generator is defined as $\bm{a}_{t,i}=[p_{t,i},v_{t,i}]$, where $p_{t,i}$ and $v_{t,i}$ are its active power and voltage magnitude, respectively. The generator working in the P/Q control mode can be defined as $\bm{a}_{t,i}=[p_{t,i},q_{t,i}]$, where $q_{t,i}$ is the reactive power of the generator.

\textbf{Probability Space $\mathcal{P}$}: $\mathcal{P}$ is the probability set of transitioning to the next state $\bm{s}_{t+1,n}$ from the current state $\bm{s}_{t}$ after taking a deterministic action $\bm{a}_t$.

\textbf{Reward Space $\mathcal{R}$}: A reward $r_t$ is obtained after taking action $\bm{a}_t$ at state $\bm{s}_t$, which indicates the effectiveness of the selected action. The goal of DRL is to learn an optimal action strategy $\pi(\bm{a}_t|\bm{s}_t)$ to maximize the expected cumulative rewards. Hence, it is essential to encode the objective of the optimization problem into a reward function to facilitate the DRL training. The reward function of the MDP can be formulated as
\begin{equation}
\label{eq:reward_func}
    \begin{aligned}
        r_t &= w^\text{vol}\left\{\sum_{n=1}^{N^\text{L}} \exp\left[-\left( 1-\left|\mathbb{V}_{t,n}^\text{L}\right| \right)^2\right]\right\}^{\frac{1}{2}},\\
        &+ w^\text{RER}\left[\sum_{j=1}^{N^\text{W}} \exp\left(\frac{p_{t,j}^\text{W,act}}{\bar{p}_{t,j}^\text{W}}\right) + \sum_{k=1}^{N^\text{S}}\exp\left(\frac{p_{t,k}^\text{S,act}}{\bar{p}_{t,k}^\text{S}}\right)\right],\\
        &+w^\text{gen}\left\{\sum_{i=1}^{N^\text{T}}\exp\left(-C_{t,i}^\text{T}\right) +\sum_{j=1}^{N^\text{W}} \exp\left[-\left(C^\text{W,r}_{t,j}+C^\text{W,p}_{t,j}\right)\right]\right.\\
        &\left.\qquad\qquad\qquad\qquad +\sum_{k=1}^{N^\text{S}}\exp\left[-\left(C^\text{S,r}_{t,k}+C^\text{S,p}_{t,k}\right)\right] \right\},
    \end{aligned}
\end{equation}
where the above three reward terms correspond to the objective functions of the optimization problem, i.e., voltage control $J^\text{vol}$, renewable accommodation $J^\text{RER}$, and generation cost minimization $J^\text{gen}$.

We then introduce how we handle the optimization constraints defined from Eq. \eqref{eq:cons_active_power_balance} to \eqref{eq:cons_branch_flow} and encode them as rewards into the DRL training process. For the power balance constraints in Eq. \eqref{eq:cons_active_power_balance} and \eqref{eq:cons_reactive_power_balance}, if these two constraints cannot be satisfied during the DRL training process, a constant penalty term ($-10$ in our simulation by default) will be added to the reward for constraint violation. Moreover, the current training episode will be terminated. Subsequently, the DN environment will be reset and the input MDP state for the next episode will be randomly initialized. For the load voltage constraint, generator reactive power constraints, and branch flow constraint presented in Eq. \eqref{eq:cons_load_voltage}, \eqref{eq:cons_gen_reactive_power_thermal}-\eqref{eq:cons_gen_reactive_power_solar}, and \eqref{eq:cons_branch_flow}, respectively, if these constraints are violated, the same constant penalty term ($-10$ by default) is added to the reward for violation feedback. For the generator's voltage and power constraints defined in Eq. \eqref{eq:cons_gen_voltage_thermal}-\eqref{eq:cons_gen_voltage_solar} and \eqref{eq:cons_gen_active_power_thermal}-\eqref{eq:cons_gen_active_power_solar}, respectively, their corresponding values will not violate these inequality constraints. This is because the actions of each generator are its power and voltage defined as $\bm{a}_{t,i}=[p_{t,i},v_{t,i}]$, indicating that the lower and upper limits of these constraints are encoded as the bounds of the MDP's action space. As a result, the action value of each generator will always meet these constraints.

\subsubsection{Solving MDP by DDPG} \label{subsubsec:method_DRL_DDPG}
The objective of DRL is to maximize the expected cumulative rewards, denoted by $\bar{R}_\theta$, which can be formulated as
\begin{equation}
    \label{eq:DRL_obj}
    \begin{aligned}
    \bar{R}_{\theta} &=\mathbb{E}_{\bm{a}_t\sim\pi,r_t,\bm{s}_{t+1}\sim \mathbb{P}}\left[R(\tau)\right],\\
    &= \sum_{\tau} R(\tau)P(\tau\mid \theta),
    \end{aligned}
\end{equation}
where $\tau$ represents the trajectory of the MDP transitions, recording all 4-tuple transitions denoted by $\{\bm{s}_t,\bm{a}_t,r_t,\bm{s}_{t+1}\}$ from the beginning of $\tau$ to its end, $R(\tau)$ represents the cumulative rewards of the trajectory, $\theta$ represents the parameters of our action strategy $\pi$, $P\left(\tau|\theta\right)$ is the occurrence probability of the trajectory $\tau$.

We then introduce DDPG~\cite{lillicrap2019} to maximize $\bar{R}_\theta$. DDPG is the most representative actor-critic DRL algorithm to optimize the derived MDP. The major difference between DDPG and other DRL algorithms is that the action policy $\pi(\bm{a}_t|\bm{s}_t)$ deterministically outputs the values of actions instead of the probability distribution of actions, which dramatically decreases the computation cost and makes it much easier to implement. Specifically, policy gradient method is applied in DDPG to update our action policy, which can be formulated as
\begin{equation}
    \label{eq:policy_gradient}
    \theta \gets \theta + \eta_\theta\nabla \bar{R}_\theta,
\end{equation}
with the gradient of the DRL objective defined as
\begin{equation}
    \label{eq:reward_gradient}
    \nabla \bar{R}_\theta = \frac{1}{N_\tau} \sum^{N_\tau} \sum^{T_\tau} \mathcal{A}_\theta\left(\bm{s},\bm{a}\right)\nabla \log p\left(\bm{a}|\bm{s},\theta\right),
\end{equation}
where $\eta_\theta$ is the learning rate, $N_\tau$ is the number of trajectories, $T_\tau$ is the length of each trajectory, and $\mathcal{A}_\theta\left(\bm{s},\bm{a}\right)$ represents the advantage function assessing the effectiveness of the state-action pair compared to a certain baseline, which can be formulated as
\begin{equation}
\label{eq:adv_func}
\begin{aligned}
    \mathcal{A}_\theta\left(\bm{s}_t,\bm{a}_t\right) &=  G_t -b, \\
    & =\sum_{t'=t}^{T_\tau}\gamma^{t'-t}r_{t'} - \mathbb{E}_{s\sim \mathbb{P}}\left[R(\bm{s})\right],
\end{aligned}
\end{equation}
where $\gamma$ is a discounting factor, $b$ represents the baseline of reward considering all possible actions, and $\mathbb{E}_{s\sim \mathbb{P}}\left[R(\bm{s})\right]$ is the expected cumulative rewards considering all possible states at the current time step. Due to the uncertainty of the MDP transitions, both $G_t$ and $b$ are random variables. To accurately estimate the advantage function, DDPG introduces a critic network $Q_\phi (\bm{s}_t,\bm{a}_t)$ formulated as
\begin{equation}
    \label{eq:critic_net}
    Q_\phi (\bm{s}_t,\bm{a}_t) = \mathbb{E}_{r_t,\bm{s}_{t+1}\sim \mathbb{P}}\left\{r_t+\gamma Q_\phi\left[(\bm{s}_{t+1},\pi_\theta\left(\bm{s}_{t+1}\right)\right]\right\},
\end{equation}
where $\phi$ represents parameters of the critic networks. With the critic network, the advantage function in Eq. \eqref{eq:adv_func} can be rewritten as
\begin{equation}
\label{eq:adv_func_rewrite}
\mathcal{A}_\theta\left(\bm{s}_t,\bm{a}_t\right) = Q_\phi\left(\bm{s}_t,\bm{a}_t\right)-\sum_{\bm{a}_t}Q_\phi\left(\bm{s}_t,\bm{a}_t\right).
\end{equation}

\begin{algorithm}[!t]
\caption{The DRL-based Control Strategy} 
\label{algo:ddpg}
\begin{algorithmic}
\STATE Initialize action policy $\pi_\theta(\bm{a}|\bm{s})$ and critic network $Q_\phi(\bm{s},\bm{a})$
\STATE Initialize target networks $\pi_{\theta'}$ and $Q_{\phi'}$
\STATE Initialize the replay buffer $\mathcal{B}$
\FOR{$n=1,2,\cdots,N_\tau$}
\STATE Initialize a Gaussian noise $\mathcal{N}$ for action exploration
\FOR{$t=1,2,\cdots,T_\tau$}
\STATE Receive the output of MG-ASTGCN $\bm{y}_t$ 
\STATE Prepare the input state $\bm{s}_t$
\STATE Get action $\bm{a}_t$ = $\pi_\theta(\bm{s}_t)$ + $\mathcal{N}\sim(0,0.1)$.
\STATE Get reward $r_t$ and transit to the next state $\bm{s}_{t+1}$
\STATE Store a transition $\{\bm{s}_t,\bm{a}_t,r_t,\bm{s}_{t+1}\}$ in $\mathcal{B}$
\STATE Randomly sample a batch of $N$ transitions from $\mathcal{B}$.
\STATE Calculate $q_t$ using the target networks.
\STATE Update the critic network $Q_\phi$:
\STATE $\qquad \qquad \qquad \phi \gets \phi-\eta_\phi\nabla_\phi L(Q_\phi)$
\STATE Update the action policy $\pi_\theta$:
\STATE $\qquad \qquad \qquad \theta\gets \theta+\eta_\theta\nabla_\theta\bar{R}_\theta$
\STATE Update target networks via soft update :
\STATE $\qquad \qquad \qquad \theta' \gets \rho\theta+(1-\rho)\theta'$
\STATE $\qquad \qquad \qquad \phi' \gets \rho\phi+(1-\rho)\phi'$.
\ENDFOR
\ENDFOR
\end{algorithmic}
\end{algorithm}

Providing that calculating the output value of the critic network depends on environments rather than our action policy $\pi$, it is applicable to learn the critic network $Q_\phi$ in an off-policy manner taking advantage of transitions generated from a different action policy denoted by $\pi_{\theta'}$. The critic network can be updated by minimizing the root mean square error formulated as
\begin{equation}
\label{eq:critic_net_update}
L\left(Q_\phi\right) = \mathbb{E}_{\bm{s}_t,r_t\sim\mathbb{P}, \bm{a}_t\sim \pi_\theta(\bm{s}_t)} \left\{\left[Q_\phi\left(\bm{s}_t,\bm{a}_t\right)-q_t\right]^2\right\},
\end{equation}
with the estimated critic network via the target action and critic networks $\pi_{\theta'}, Q_{\phi'}$ defined as
\begin{equation}
    q_t = r_t + \gamma Q_{\phi'}\left[\bm{s}_{t+1},\pi_{\theta'}\left(\bm{s}_{t+1}\right)\right].
\end{equation}
Similarly, the critic network is updated via gradient descent formulated as
\begin{equation}
    \phi \leftarrow \phi - \eta_\phi\nabla_\phi L(Q_\phi),
\end{equation}
where $\eta_\phi$ is the learning rate for updating the critic network.

The proposed two target networks are updated in an exponential moving average manner using parameters of their corresponding action and critic networks, which can be formulated as
\begin{align}
    \theta' &\leftarrow \rho\theta+(1-\rho)\theta',\\
    \phi' &\leftarrow \rho\phi+(1-\rho)\phi',
\end{align}
where $\rho$ is the smoothing parameter.

\begin{figure}[!t]
    \centering
    \includegraphics[width=\linewidth]{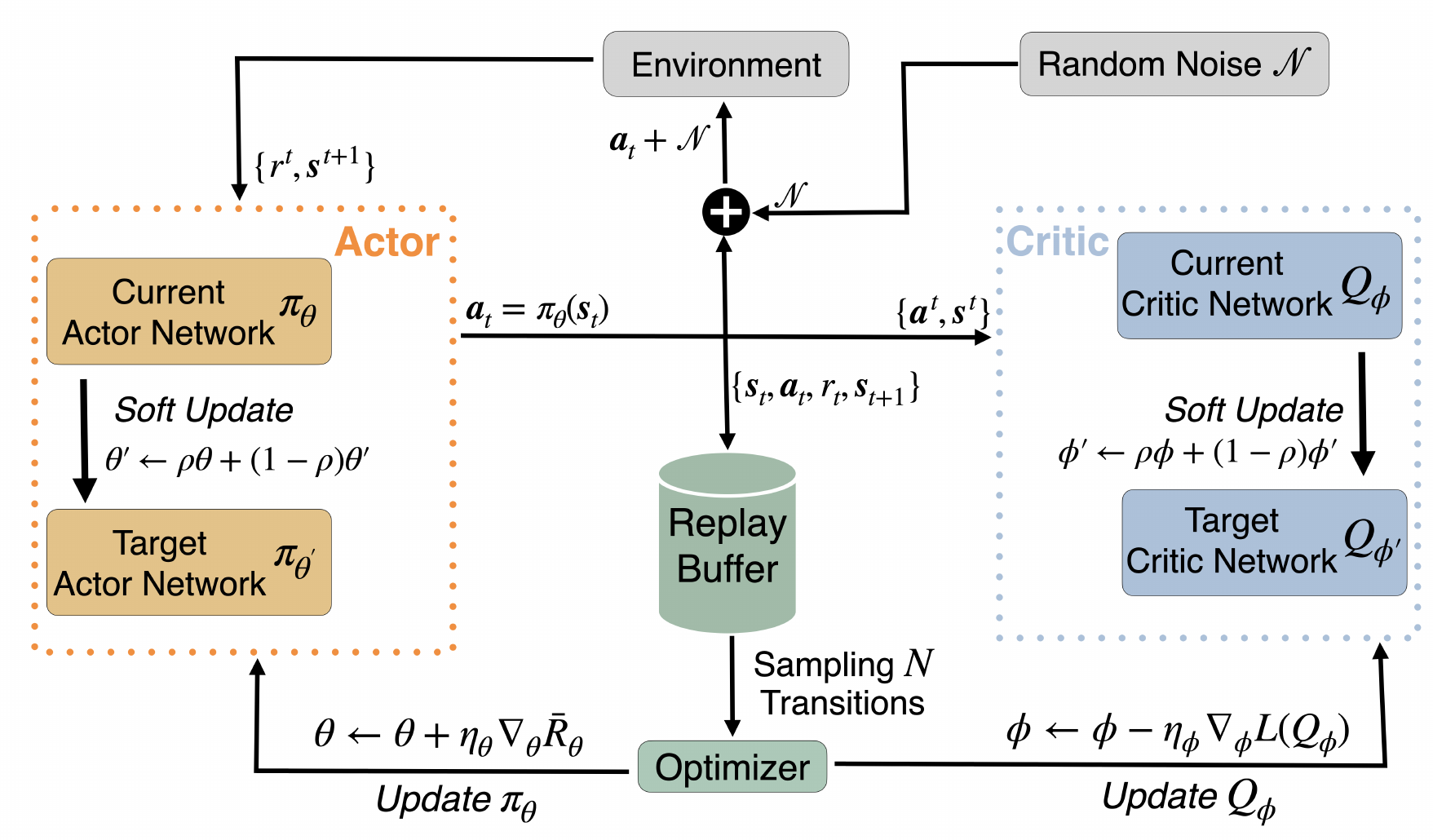}
    \caption{The workflow of DDPG.}
    \label{fig:ddpg_workflow}
\end{figure}

In summary, DDPG follows the policy-gradient way to optimize the derived MDP, where a critic network is introduced to assess the learned action policy. The workflow illustration and algorithmic procedure of the DDPG are presented in Fig. \ref{fig:ddpg_workflow} and Algorithm \ref{algo:ddpg}, respectively.

\begin{table}[!t]
    \centering
    \caption{System characteristics of IEEE 33, 69, and 118-bus RDSs.}
    \begin{tabular}{ M{4.2cm}  M {0.9cm}  M{0.9cm}  M{1.1cm} }
    \hline
    \textbf{RDSs Charateristics}  &  $\bm{33}$\textbf{-Bus}    & $\bm{69}$\textbf{-Bus}  & $\bm{118}$\textbf{-Bus}\\
    \hline
    Total Buses ($N^\text{L}$)  &   $33$  &   $69$  &   $118$\\[0.5ex]
    Thermoelectric Generators ($N^\text{T}$)   &   $2$   &   $3$   &$4$\\[0.5ex]
    Wind Turbines ($N^\text{W}$) &   $5$   &   $10$  &   $15$\\[0.5ex]
    Solar PV Generators ($N^\text{S}$) &   $5$   &   $10$  &   $15$\\[0.5ex]
    Baseline Voltage (kV)   &   $12.66$ &   $12.66$ &   $11$\\[0.5ex]
    Baseline Apparent Power (MVA)   &   $100$   &   $100$   &   $100$\\[0.5ex]
    Total Load Active Power (MW)    &   $3.715$  &   $3.800 $  &   $22.710$\\[0.5ex]
    Total Load Reactive Power (MVAR)    &   $2.300$   &   $2.690$   &   $17.041$\\[0.5ex]
    \hline
    \end{tabular}
    \label{tab:ieee_model}
\end{table}

\begin{table}[!t]
    \centering
    \caption{The initialized parameters.}
    \begin{tabular}{cc||cc}
    \hline
    $a_i$ & $0.0175$ & $b_i$ & $1.75$\\
    $c_i$ & $0$ & $c_j^\text{W,r},c_k^\text{S,r}$ & $1.5$\\
    $c_j^\text{W,p},c_k^\text{S,p}$ & $3$ & $w^\text{vol}$ & $1$\\
    $w^\text{RER}$ & $1$ & $w^\text{gen}$ & $0.01$\\
    $T^\text{r}$ & $32$ & $T^\text{d}$ & $16$\\
    $T^\text{w}$ & $4$ & $N^\text{d}$ & $24$\\
    $\gamma$ & $0.99$ & $\rho$ & $0.01$\\
    $\eta_\theta$ & $0.0003$ & $\eta_\phi$ & $0.0003$\\
    \hline
    \end{tabular}
    \label{tab:parameters}
\end{table}

\begin{figure}[!t]
    \centering
    \subfloat[Power Output Profiles.]{
    \includegraphics[width=0.47\linewidth]{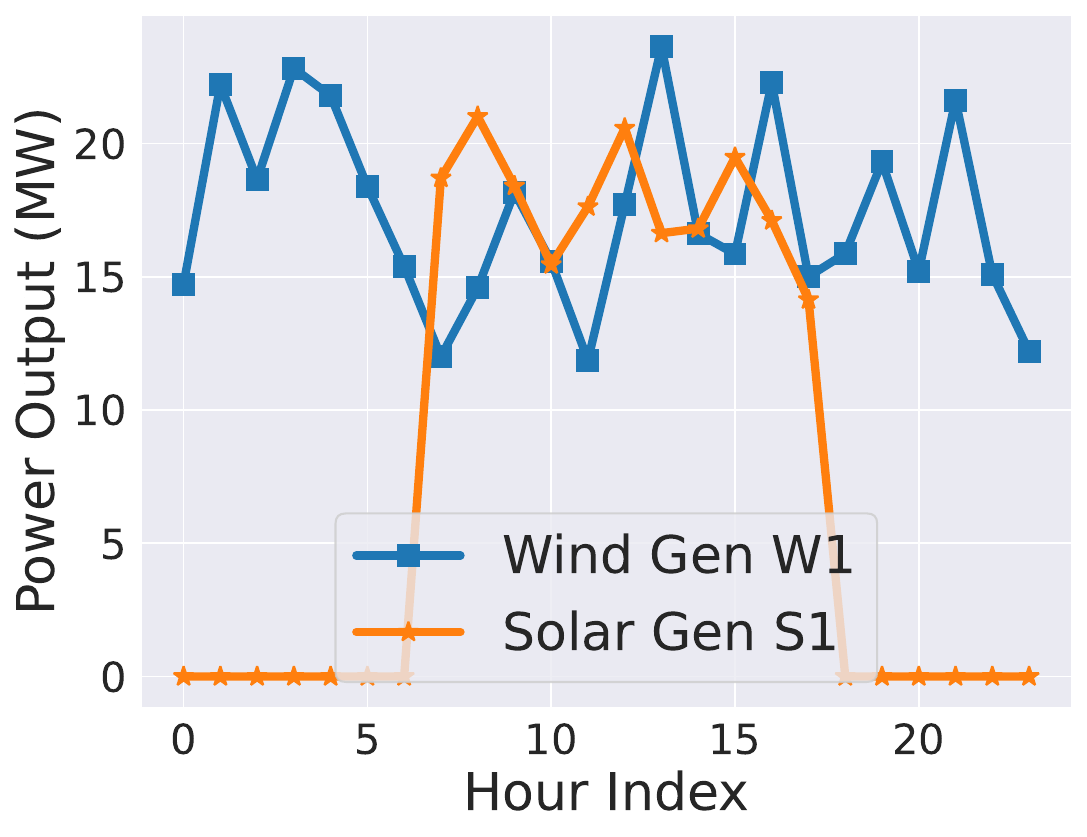}
    \label{fig:power_profile}
    }
    \subfloat[Part of 118-Bus System.]{
    \includegraphics[width=0.50\linewidth]{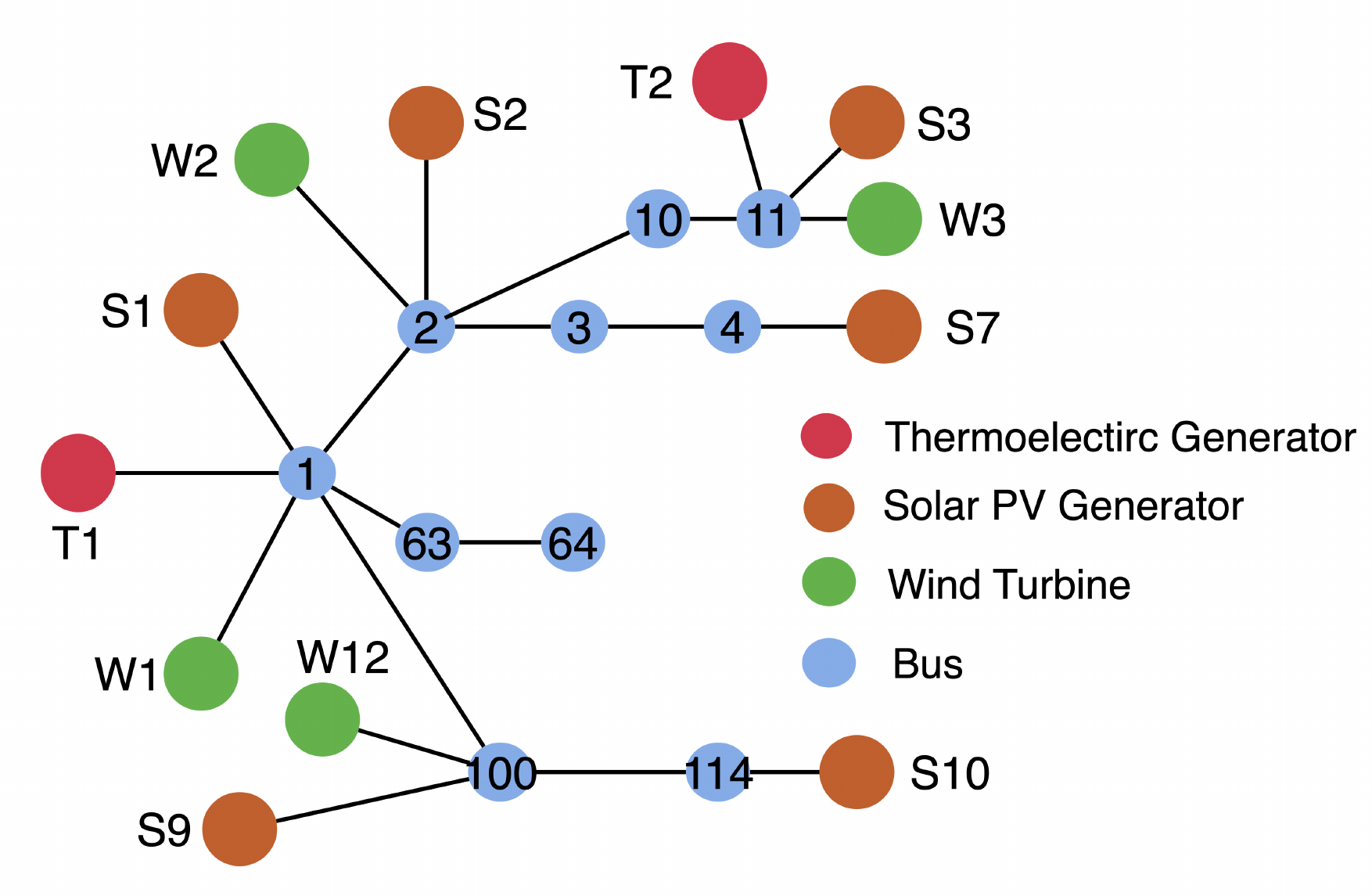}
    \label{fig:power_profile_test_system}
    }
    \caption{The power profiles of one wind generator and one solar generator throughout one day in a part of the IEEE 118-bus test system.}
    \label{fig:power_profile_wind_solar_gen}
    \vspace{-3mm}
\end{figure}

\section{Experiments and Results} \label{sec:exps}
\subsection{Experimental Settings} \label{subsec:exps_settings}
\subsubsection{Evaluation Scenario} \label{subsubsec:exps_settings_scenario}
The proposed DRL-based strategy is tested on the modified IEEE 33-bus, 69-bus, and 118-bus radial distribution systems (RDSs)~\cite{schneider2018} with their detailed statistics provided in Table \ref{tab:ieee_model}. The time interval between two consecutively operational time steps is one hour, aligning with previous studies in DN's voltage control~\cite{cao2020,cao2020-2,cao2021,cao2022}. The simulation lasts for a one-year duration to ensure sufficient DRL training. One Nvidia TITAN RTX graphics processing unit is used for DRL training. The batch size of the DDPG algorithm is set as $256$. The random noise employed in the DDPG is a Gaussian noise, denoted by $\mathcal{N}\sim(\mu,\sigma)$ with its mean $\mu$ and standard deviation $\sigma$ setting as $0$ and $0.1$, respectively. The number of hidden layers for the actor and critic networks is $3$, each of which consists of $400$ neurons. We adopt the Adam optimizer to update our actor, i.e., $\pi_\theta$ and critic, i.e., $Q_\phi$, networks. The number of ST components inside our MG-ASTGCN is $3$. The initialized parameters of our DRL-based control strategy are provided in Table \ref{tab:parameters}. 

Note that we use the same reserve cost coefficients and penalty cost coefficients for wind and solar generators provided by~\cite{khan2020}, i.e., $c_j^\text{W,r}=c_k^\text{S,r}=1.5$ and $c_j^\text{W,p}=c_k^\text{S,p}=3$. For the cost coefficients of the thermoelectric generator, we use the same coefficients provided by~\cite{panda2015}, i.e., $a_i=0.0175$, $b_i=1.75$, and $c_i=0$. Additionally, we provide the power output profile of one wind power generator and one solar generator throughout one day in a part of the IEEE $118$-bus test system in Fig. \ref{fig:power_profile_wind_solar_gen}.

\subsubsection{Algorithm Performance Metric}
In our experiments, the reward function defined in Eq. \eqref{eq:reward_func} is adopted to measure the performance of both DDPG and benchmark algorithms. Specifically, we introduce three metrics, namely SCORE, voltage fluctuation rate denoted by $\alpha^\text{vol}$, and RER accommodation rate denoted by $\alpha^\text{RER}$, to examine the effectiveness of the learned control strategy, voltage control, and renewable accommodation, respectively, which are defined as
\begin{align}
\label{eq:SCORE}
    \text{SCORE} &= \frac{1}{N^{\text{eval}}}\sum_{n=1}^{N^{\text{eval}}}\sum_{t=1}^{T^{\text{end}}} r_t,\\
    \alpha^\text{vol} &= \frac{1}{N^\text{eval}T^\text{end}}\sum_{n=1}^{N^{\text{eval}}}\sum_{t=1}^{T^{\text{end}}} J_t^\text{vol} \times 100\%,
\end{align}
\begin{align}
    \alpha^\text{RER} &= \frac{1}{N^\text{eval}T^\text{end}}\frac{1}{N^\text{W}+N^\text{S}}\sum_{n=1}^{N^{\text{eval}}}\sum_{t=1}^{T^{\text{end}}} J_t^\text{RER} \times 100\%,
\end{align}
where $N^{\text{eval}}$ is the number of episodes for evaluation and $T^{\text{end}}$ is the length of each episode. Both $N_{\text{eval}}$ and $T_{\text{end}}$ are initialized as $100$.

\begin{table*}[!t]
    \centering
    \caption{The evaluation results of four benchmarks, i.e., HHO, GWO, IP, and LQP, and our DRL-based method.}
    \begin{tabular}{c|c|c|c|c|c|c}
    \multicolumn{2}{c|}{} & HHO & GWO & IP & LQP & \textbf{Ours}\\
    \hline
    \multirow{4}{*}{33-Bus} & Time Cost per Step (secs) & $2.4$ & $3.4$ & $2.1$ & $3.2$ & $\bm{0.9}$\\
    \cline{2-7}
    & SCORE & $3531$ & $3014$ & $3681$ & $3778$ & $\bm{4015}$\\
    \cline{2-7}
    & Voltage Fluctuation Rate & $1.52\%$ & $2.36\%$ & $0.95\%$ & $0.58\%$ & $\bm{0.22\%}$\\
    \cline{2-7}
    & RER Accommodation Rate & $78.9\%$ & $77.1\%$ & $86.2\%$ & $84.4\%$ & $\bm{94.2\%}$\\
    \hline
    \multirow{4}{*}{69-Bus} & Time Cost per Step (secs) & $6.1$ & $8.1$ & $5.5$ & $9.2$ & $\bm{1.3}$\\
    \cline{2-7}
    & SCORE & $6105$ & $5822$ & $6716$ & $6704$ & $\bm{7757}$\\
    \cline{2-7}
    & Voltage Fluctuation Rate & $1.98\%$ & $2.59\%$ & $0.69\%$ & $0.85\%$ & $\bm{0.19\%}$\\
    \cline{2-7}
    & RER Accommodation Rate & $80.5\%$ & $82.8\%$ & $84.5\%$ & $89.2\%$ & $\bm{93.4\%}$\\
    \hline
    \multirow{4}{*}{118-Bus} & Time Cost per Step (secs) & $12.7$ & $15.8$ & $9.2$ & $17.0$ & $\bm{1.6}$\\
    \cline{2-7}
    & SCORE & $8232$ & $7481$ & $11614$ & $12171$ & $\bm{14384}$\\
    \cline{2-7}
    & Voltage Fluctuation Rate & $2.48\%$ & $3.04\%$ & $0.58\%$ & $0.52\%$ & $\bm{0.20\%}$\\
    \cline{2-7}
    & RER Accommodation Rate & $82.9\%$ & $77.6\%$ & $90.3\%$ & $86.4\%$ & $\bm{94.5\%}$\\
    \hline
    \end{tabular}
    \label{tab:eval_res}
\end{table*}

\begin{figure*}[!t]
\centering
\subfloat[33-bus]{
\includegraphics[width=0.3\linewidth]{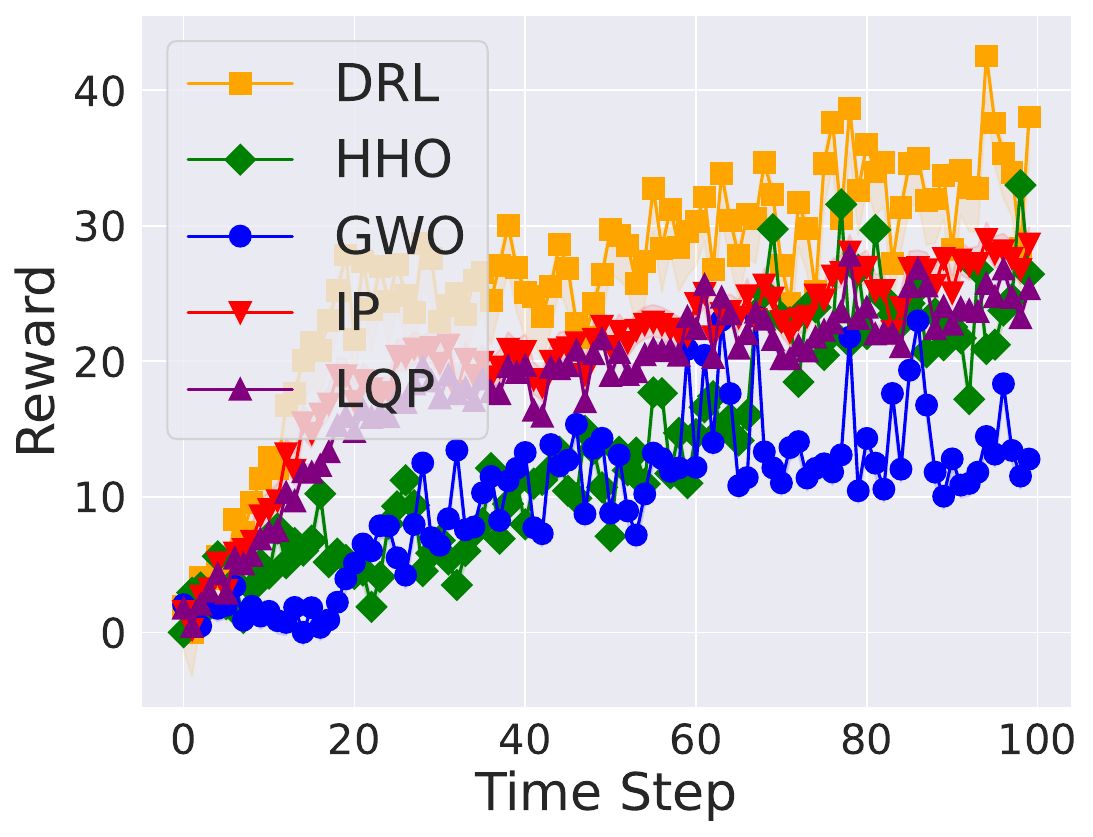}
\label{fig:eval_res_33bus}
}
\hspace{0.4em}
\subfloat[69-bus]{
\includegraphics[width=0.3\linewidth]{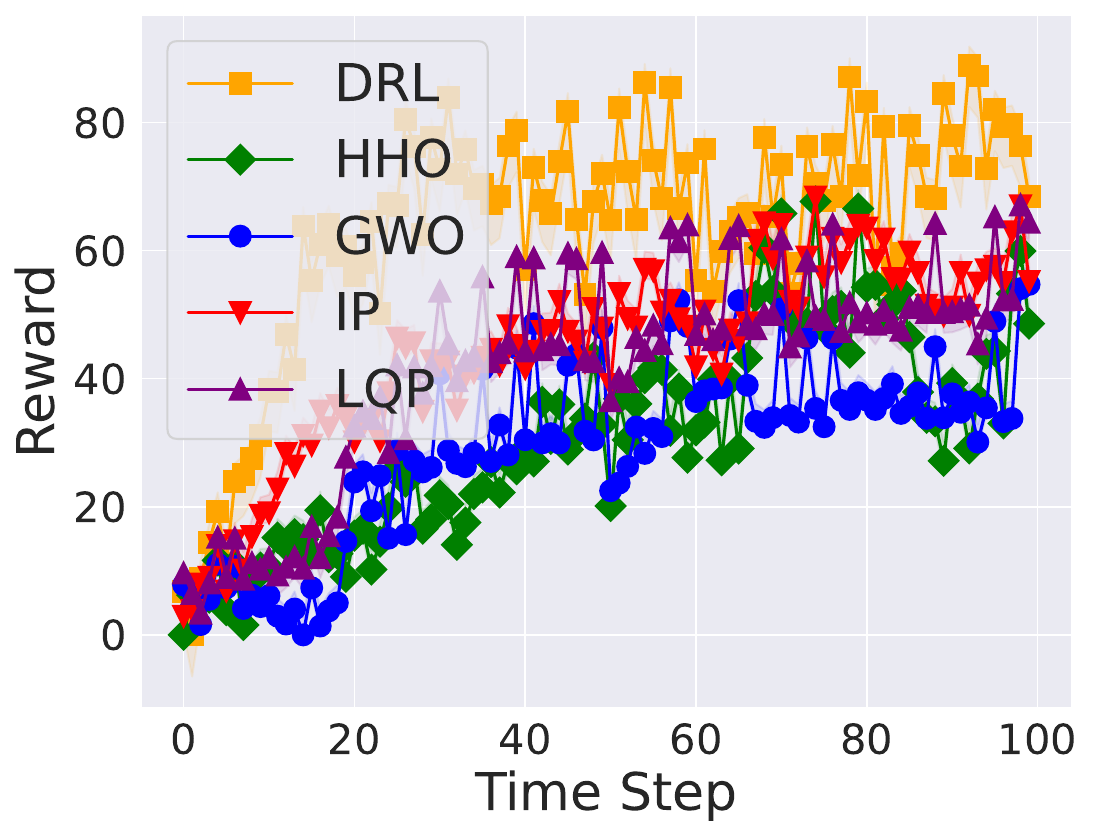}
\label{fig:eval_res_69bus}
}
\hspace{0.4em}
\subfloat[118-bus]{
\includegraphics[width=0.3\linewidth]{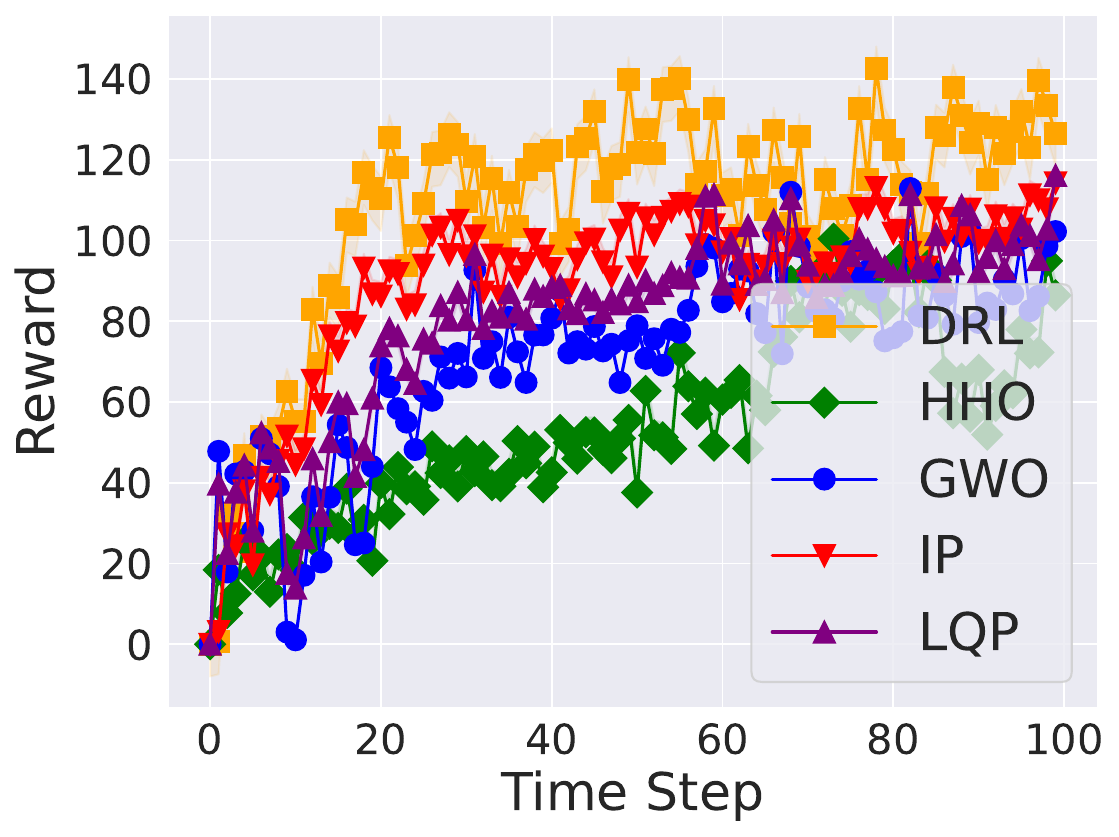}
\label{fig:eval_res_118}
}
\caption{The evaluation rewards of HHO, GWO, IP, LQP, and DRL-based strategy on three RDSs in the first $100$ time steps.}
\label{fig:eval_res}
\end{figure*}

\subsection{Experimental Results} \label{subsec:exps_results}
\subsubsection{Optimization-based Benchmark Comparisons} \label{subsubsec:exps_results_benchmark}
Two representative optimization-based algorithms---HHO~\cite{mahmoud2020_hho} and GWO~\cite{routray2020_gwo}, are adopted to compare with our DRL-based strategy. Benefiting from their meta-heuristic characteristic, both the HHO and GWO algorithms can approach the optimal control strategy without modifying or relaxing the optimization formulation. Moreover, the meta-heuristic-based methods also present good convergence and robustness abilities in the previous studies~\cite{khan2020,dokeroglu2021,hao2021}. Moreover, we also introduced two relaxation-based optimization methods, namely interior-point (IP)-based method~\cite{capitanescu2007} and linear/quadratic programming (LQP)-based method~\cite{fortenbacher2019} as benchmarks. The evaluation results derived by these methods on IEEE 33, 69, and 118-bus RDSs are illustrated in Fig. \ref{fig:eval_res} with associated statistics presented in Table \ref{tab:eval_res}. The results reveal that our proposed DRL-based strategy outperforms all benchmarks by significant margins in terms of faster testing time, higher SCOREs, lower voltage fluctuation rates, and more renewable accommodation. The reason for such performance gaps is twofold:
\begin{itemize}
    \item The optimization-based benchmarks do not need training, resulting in longer computation time for evaluation at each time step. Meanwhile, since they tend to get stuck in local optimums, their output actions are less likely to be the optimal ones, resulting in smaller rewards.
    \item Tremendous data collected by the DRL-based strategy contributes to its more effective and efficient searching in the action space. Therefore, the DRL reacts much faster at the beginning of evaluation and gradually obtains a higher reward, especially in large-scale DNs.
\end{itemize}

\begin{table}[!t]
    \centering
    \caption{The voltage fluctuation rates and renewable accommodation rates using different weight coefficients of the objective functions.}
    \begin{tabular}{c|c|c|c|c|c}
    \hline
    & $w^\text{vol}$ & $w^\text{RER}$ & $w^\text{gen}$ & \makecell{Voltage \\ Fluctuation} & \makecell{Renewable \\ Accommodation} \\
    \hline
    \multirow{5}{*}{\makecell{33 \\ Bus}} & $\textbf{1}$ & $\textbf{1}$ & $\textbf{0.01}$ & $\bm{0.22\%}$ & $\bm{94.2\%}$\\
    \cline{2-6}
    & $1$ & $1$ & $0.005$ & $0.89\%$ & $92.1\%$\\
    \cline{2-6}
    & $1$ & $1$ & $0.05$ & $2.05\%$ & $88.4\%$\\
    \cline{2-6}
    & $0.5$ & $0.5$ & $0.01$ & $1.98\%$ & $84.9\%$\\
    \cline{2-6}
    & $5$ & $5$ & $0.01$ & $0.41\%$ & $85.2\%$\\
    \hline
    \multirow{5}{*}{\makecell{69 \\ Bus}} & $\textbf{1}$ & $\textbf{1}$ & $\textbf{0.01}$ & $\bm{0.19\%}$ & $\bm{93.4\%}$\\
    \cline{2-6}
    & $1$ & $1$ & $0.005$ & $1.14\%$ & $88.5\%$\\
    \cline{2-6}
    & $1$ & $1$ & $0.05$ & $0.67\%$ & $85.2\%$\\
    \cline{2-6}
    & $0.5$ & $0.5$ & $0.01$ & $2.45\%$ & $89.2\%$\\
    \cline{2-6}
    & $5$ & $5$ & $0.01$ & $0.39\%$ & $81.9\%$\\
    \hline
    \multirow{6}{*}{\makecell{118 \\ Bus}} & $\textbf{1}$ & $\textbf{1}$ & $\textbf{0.01}$ & $\bm{0.20\%}$ & $\bm{94.5\%}$\\
    \cline{2-6}
    & $1$ & $1$ & $0.005$ & $0.41\%$ & $85.2\%$\\
    \cline{2-6}
    & $1$ & $1$ & $0.05$ & $0.89\%$ & $90.1\%$\\
    \cline{2-6}
    & $0.5$ & $0.5$ & $0.01$ & $2.23\%$ & $87.6\%$\\
    \cline{2-6}
    & $5$ & $5$ & $0.01$ & $0.48\%$ & $82.9\%$\\
    \hline
    \end{tabular}
    \label{tab:weight_searching}
\end{table}

\begin{table}[!t]
    \centering
    \caption{The average training time cost per episode of ACER, A2C, PPO, SAC, and DDPG.}
    \begin{tabular}{c||c||c||c}
    \hline
    & 33-Bus & 69-Bus & 118-Bus\\
    \hline
    A2C  & $31$ secs & $78$ secs & $105$ secs \\
    ACER  & $39$ secs & $95$ secs & $136$ secs \\
    PPO  & $36$ secs & $80$ secs & $106$ secs \\
    SAC   & $38$ secs & $86$ secs & $126$ secs \\
    \textbf{DDPG} & $\bm{30}$ secs & $\bm{74}$ secs & $\bm{102}$ secs \\
    \hline
    \end{tabular}
    \label{tab:drl_time_cost}
\end{table}

Additionally, we have conducted parameter searching for the weight coefficients of each objective function, i.e., $w^\text{vol}$, $w^\text{RER}$, and $w^\text{gen}$, to determine their optimal values, We trained and evaluated our DRL-based strategy with different weight combinations. The associated voltage fluctuation rates and renewable accommodation rates are presented in Table \ref{tab:weight_searching}. The reason we always set the same weight for voltage control and renewable accommodation is that we assume that the two objectives are of the same significance in our simulation. The parameter searching results demonstrate the effectiveness of our current weight setting, i.e., $w^\text{vol}=1$, $w^\text{RER}=1$, and $w^\text{gen}=0.01$, since it achieves the best performance in terms of lower voltage fluctuation rate and higher renewable accommodation rate.

\begin{figure*}[!t]
    \centering
    \subfloat[33-Bus]{
    \includegraphics[width=0.3\linewidth]{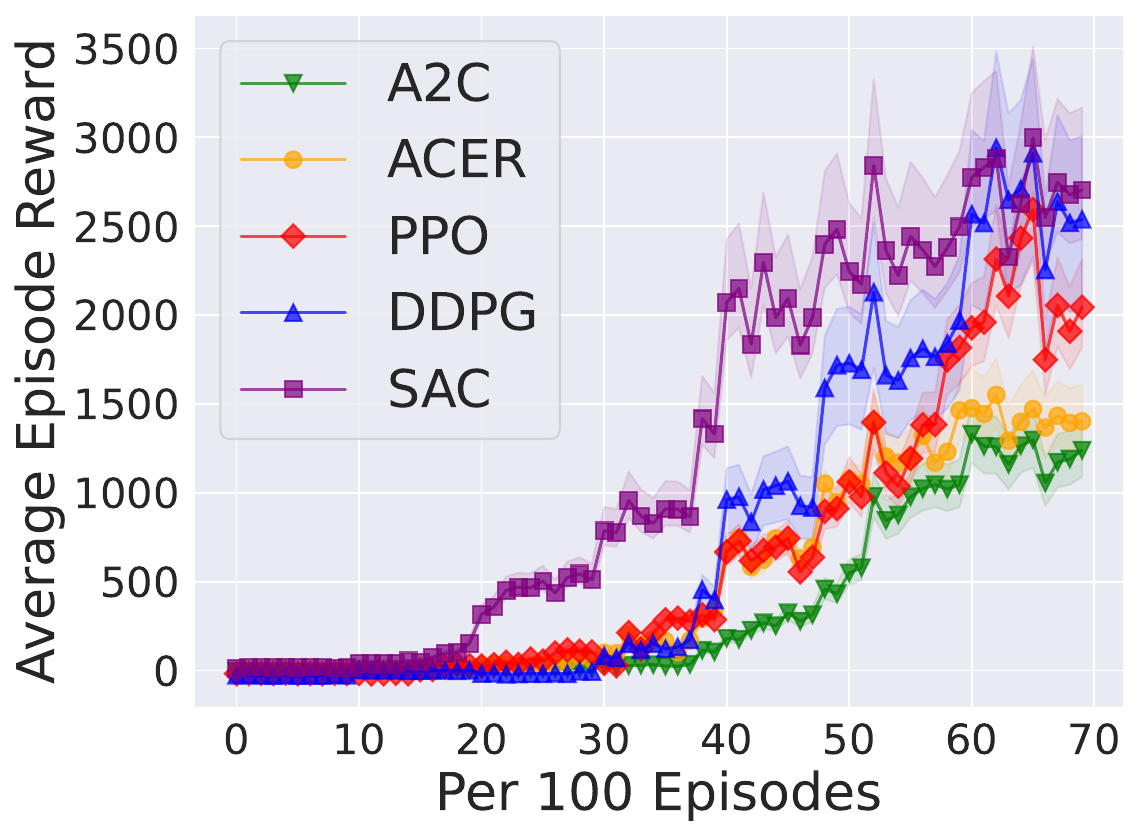}
    \label{fig:DRL_comparison_33bus}
    }
    \subfloat[69-Bus]{
    \includegraphics[width=0.3\linewidth]{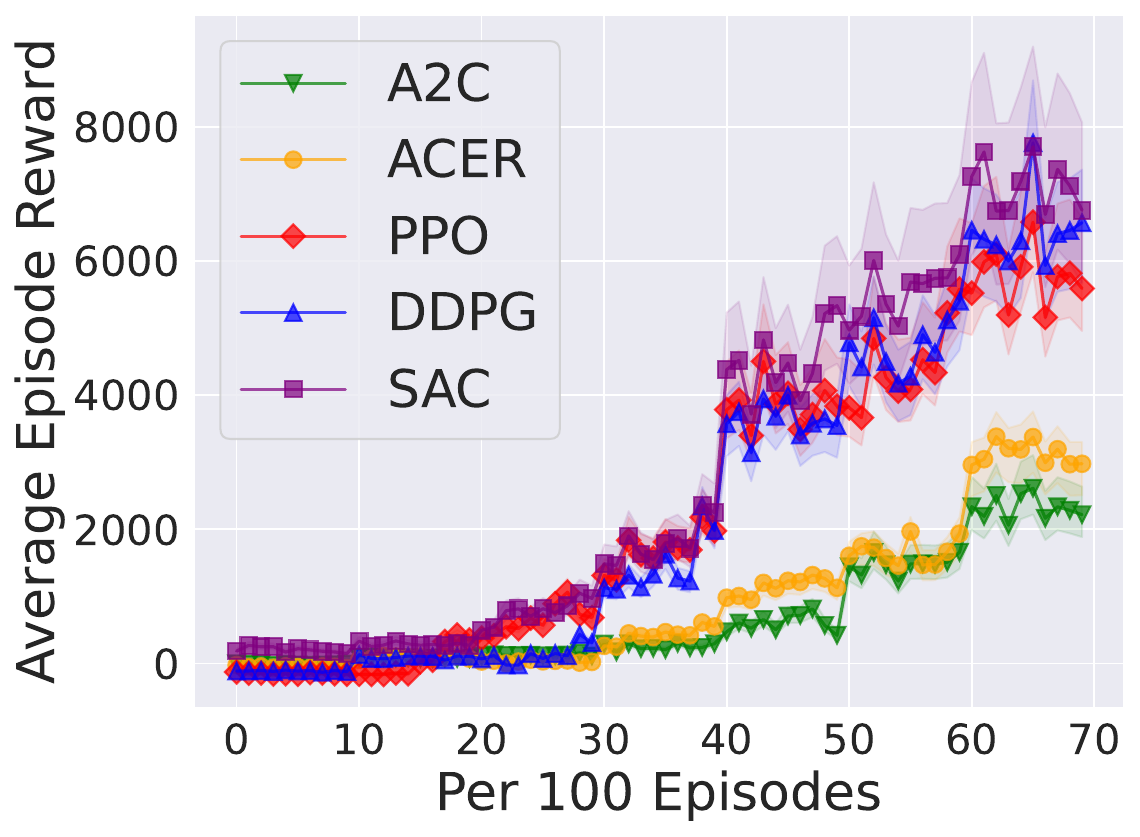}
    \label{fig:DRL_comparison_69bus}
    }
    \subfloat[118-Bus]{
    \includegraphics[width=0.3\linewidth]{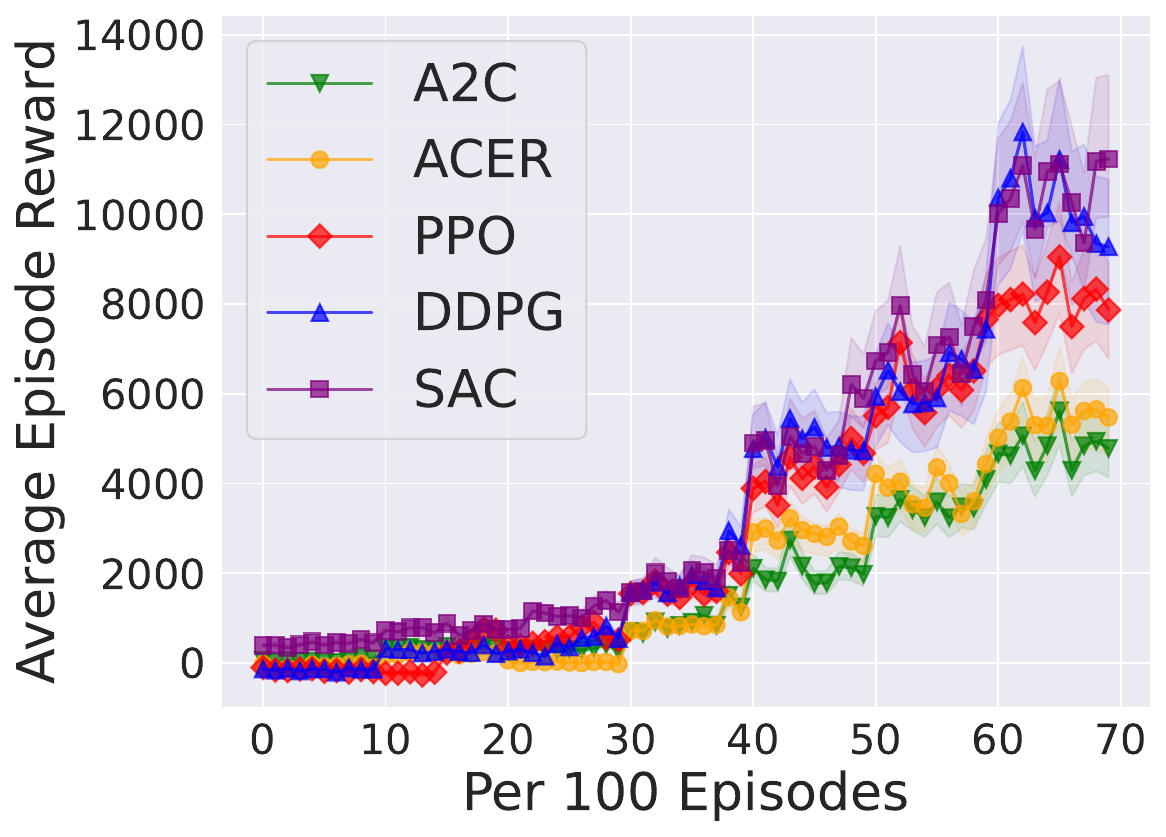}
    \label{fig:DRL_comparison_118bus}
    }
    \caption{The average episode rewards for each 100 episode during training of ACER, A2C, PPO, SAC, and DDPG algorithms}
    \label{fig:DRL_comparison}
\end{figure*}

\begin{figure*}[!t]
\centering
\subfloat[33-bus]{
\includegraphics[width=0.3\linewidth]{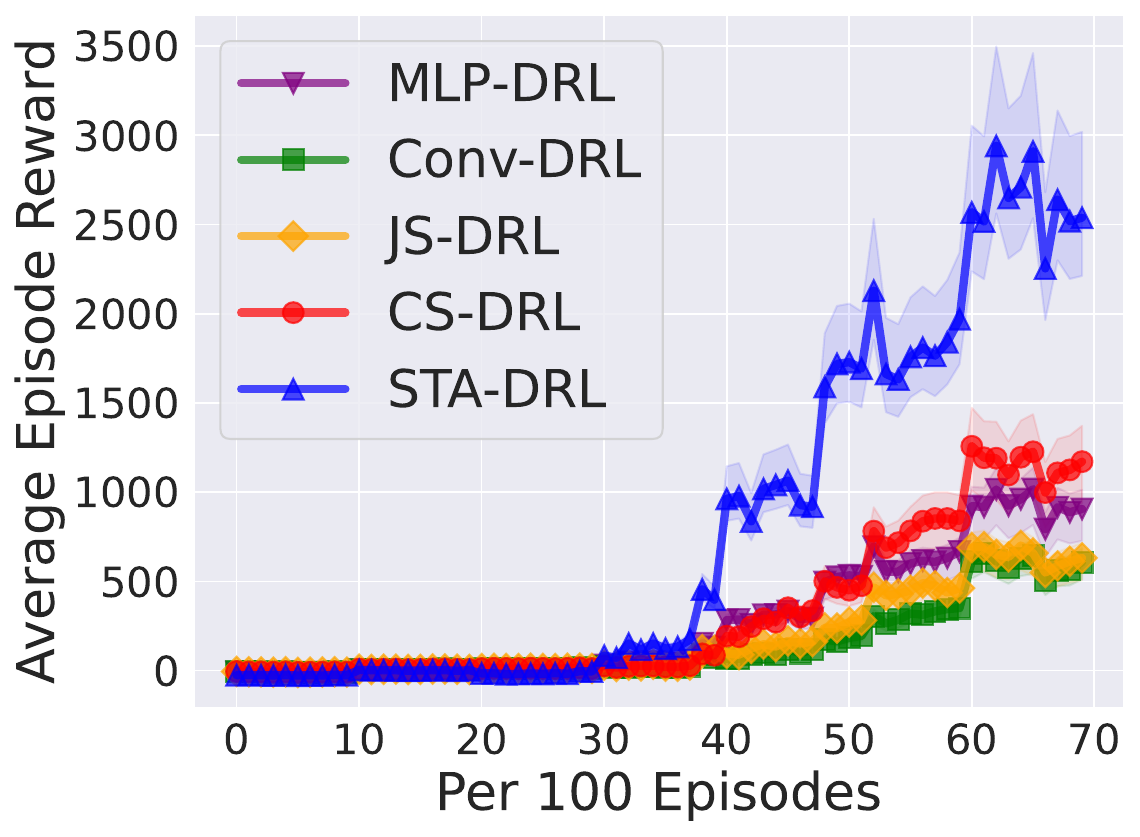}
\label{fig:att_train_33bus}
}
\subfloat[69-bus]{
\includegraphics[width=0.3\linewidth]{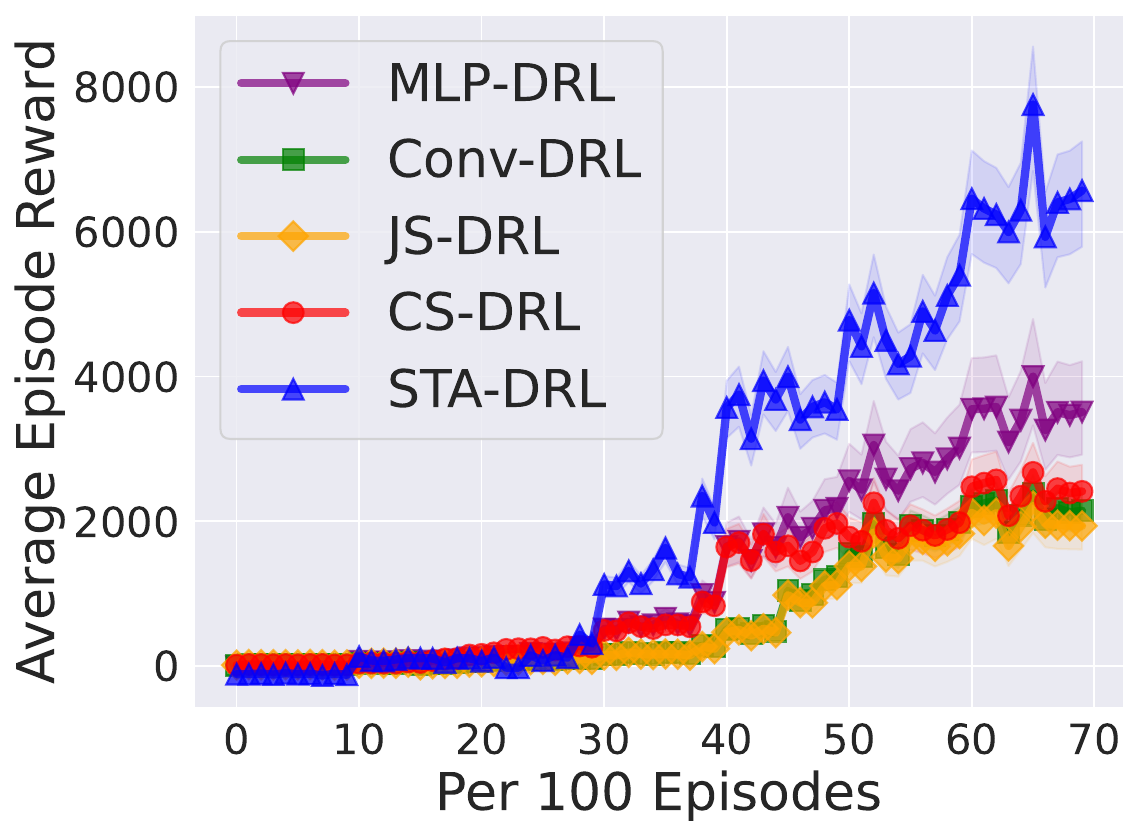}
\label{fig:att_train_69bus}
}
\subfloat[118-bus]{
\includegraphics[width=0.3\linewidth]{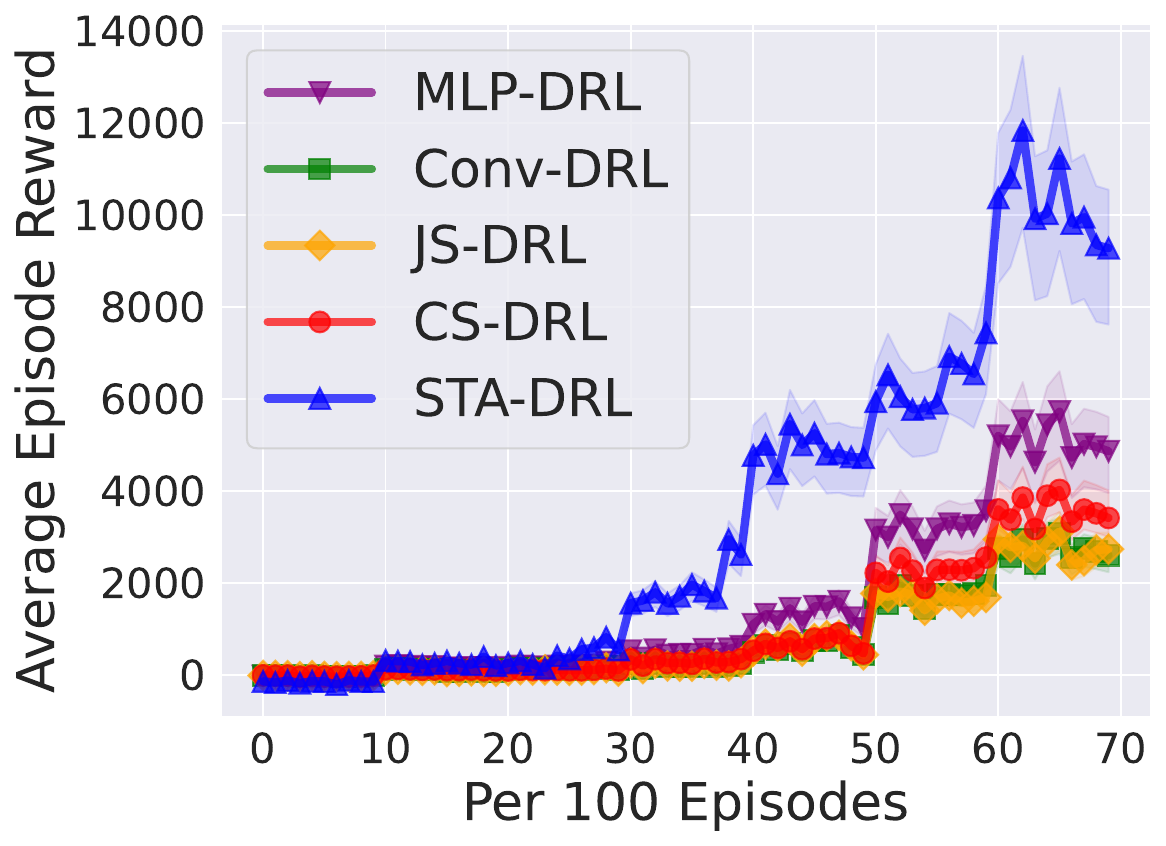}
\label{fig:att_train_118bus}
}
\caption{The average episode rewards for each $100$ episode during training of MLP-DRL, Conv-DRL, JS-DRL, CS-DRL, and STA-DRL.}
\label{fig:att_train_res}
\end{figure*}

\begin{figure*}[!t]
\centering
\subfloat[33-bus]{
\includegraphics[width=0.3\linewidth]{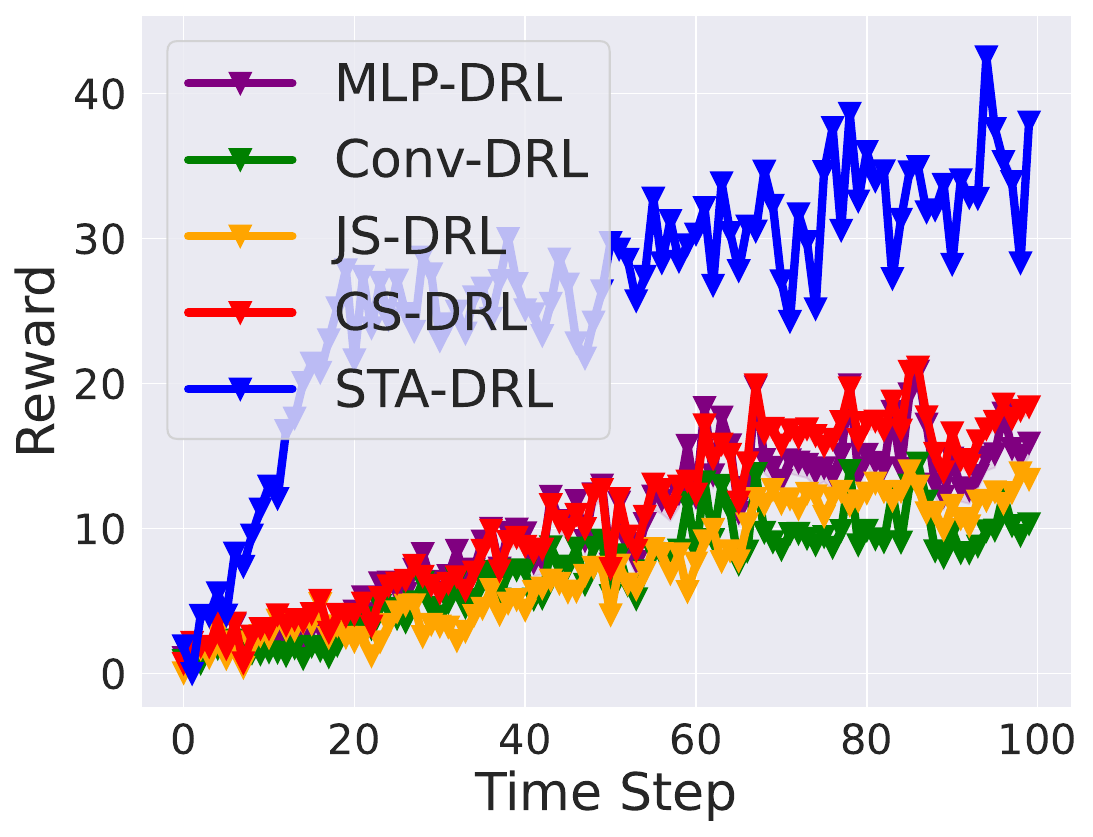}
\label{fig:att_eval_33bus}
}
\subfloat[69-bus]{
\includegraphics[width=0.3\linewidth]{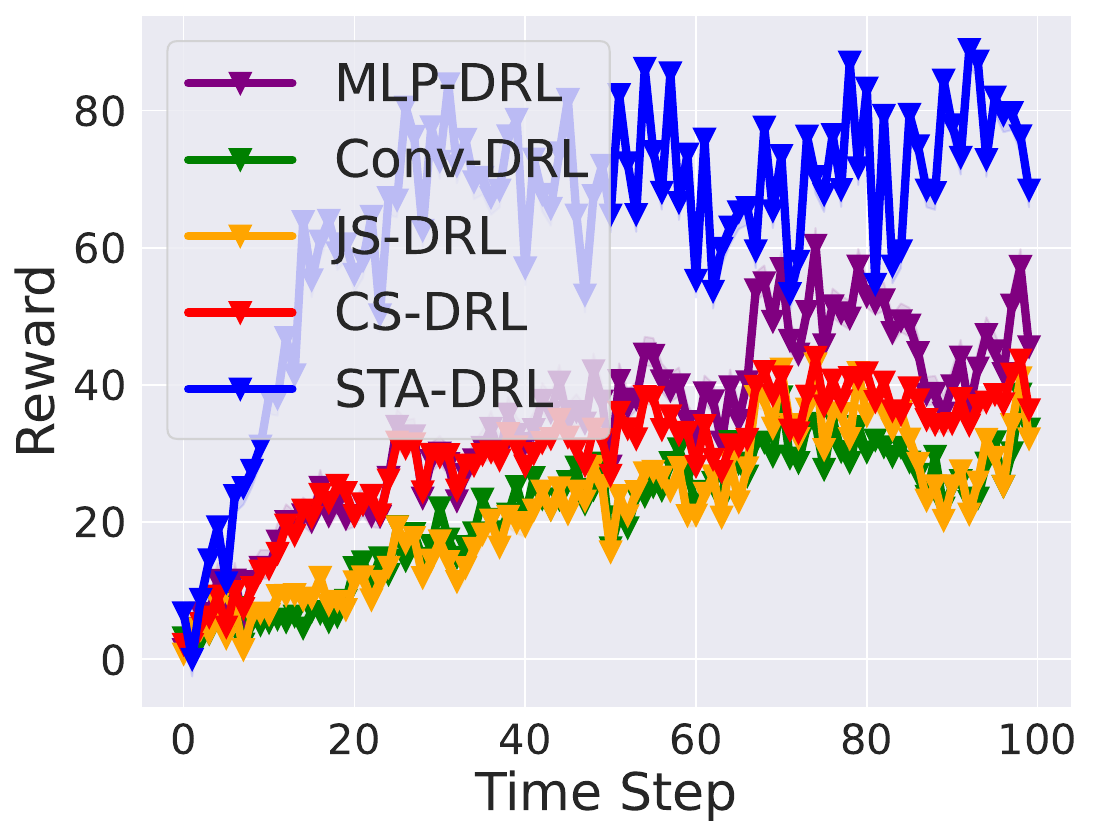}
\label{fig:att_eval_69bus}
}
\subfloat[118-bus]{
\includegraphics[width=0.3\linewidth]{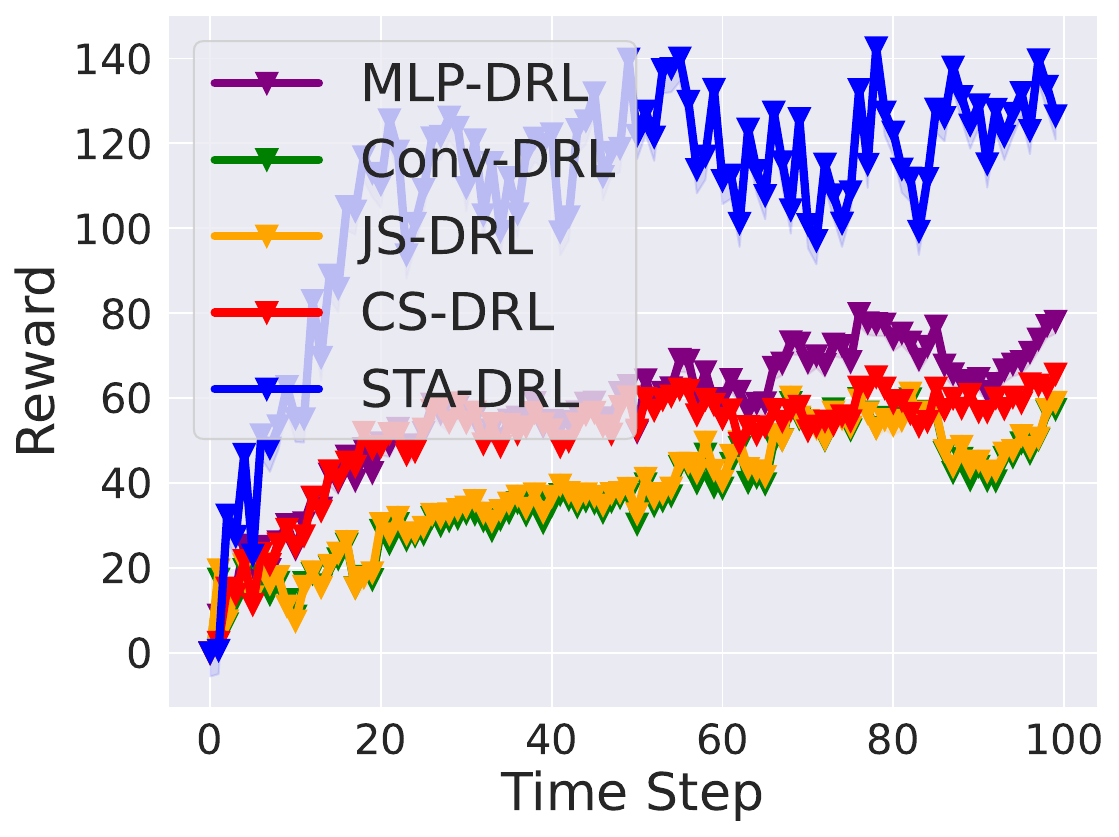}
\label{fig:att_eval_118bus}
}
\caption{The evaluation rewards of MLP-DRL, Conv-DRL, JS-DRL, CS-DRL, and STA-DRL on three RDSs in the first $100$ time steps.}
\label{fig:att_eval_res}
\end{figure*}

\subsubsection{DRL-based Baseline Comparisons} \label{subsubsec:exps_result_baseline_comp_DRL}
We also conducted comparisons of other state-of-the-art DRL algorithms, including the PPO~\cite{schulman2017_ppo} and SAC~\cite{haarnoja2018_sac}. Moreover, we have also added two more baselines -- actor-critic with experience replay (ACER)~\cite{wang2017_acer} and advantage actor-critic (A2C)~\cite{mnih2016_a2c}. The training reward curves derived from these DRL algorithms are depicted in Fig. \ref{fig:DRL_comparison}. The results show that the SAC algorithm performs slightly better than our DDPG algorithm, while the performances of the DDPG and the PPO algorithms are quite close. The reason why we still choose the DDPG algorithm is twofold: a) DDPG presents better in lower training time cost compared to the PPO and SAC algorithms, as presented in Table \ref{tab:drl_time_cost}. Specifically, the DDPG algorithm has the lowest average training time cost per episode ($30$, $74$, $102$ seconds for three respective test systems), outperforming the PPO and SAC algorithms by significant margins; b) the DDPG is relatively easy to implement with fewer hyper-parameters to tune.

\subsubsection{Effectiveness of ST Attention and ST Graphical Information of the DN} \label{subsubsec:exps_results_ST_attention}
To evaluate the effectiveness of the ST attention and its impact on the sequential DDPG, the ST attention mechanism is substituted with several other correlation extraction techniques, including cosine similarity (CS) and Jaccard similarity (JS). Their corresponding DRL strategies are termed ``CS-DRL'' and ``JS-DRL'', respectively. We term our method with ST attention as ``STA-DRL''. Moreover, we also design a baseline without the ST attention, i.e., only keeping the ST convolution module in our proposed MG-ASTGCN, which is termed ``Conv-DRL''. Furthermore, to fully verify the extracted graphical information, we additionally introduce a multilayer-perceptron-based (MLP-based) network structure for the DDPG algorithm, which is unable to extract the ST graphical information of the DN. Each layer of the MLP network is a fully-connected neural network layer. We term such a method as ``MLP-DRL''. The associated training and evaluation results of the aforementioned methods are depicted in Fig. \ref{fig:att_train_res} and \ref{fig:att_eval_res}, respectively. Note that each data point in Fig. \ref{fig:att_train_res} is calculated by taking an average of $100$ episodes' rewards. For each episode, the maximum time step is $128$, and the episode reward is the summation of each step's reward.

We can summarize several noteworthy observations regarding the effectiveness of ST attention and the extracted graphical information based on the simulation results as follows.
\begin{itemize}
    \item The outstanding training performance of our STA-DRL method indicates that the adoption of ST attention can dramatically increase DDPG's convergence speed, leading to a more effective search in the underlying action space based on the extracted ST information. As a result, our STA-DRL method significantly outperforms all baselines during evaluation as shown in Fig. \ref{fig:att_eval_res}.
    \item The alternative correlation extraction methods, i.e., CS-DRL and JS-DRL, are less effective than the ST attention, since they only focus on degree correlations among different nodes, ignoring node inner features and their temporal dependencies. Interestingly, the performance of the MLP-DRL method surpasses the CS-DRL, JS-DRL, and Conv-DRL, suggesting that adopting effective and suitable graphical information extraction, e.g., our proposed MG-ASTGCN, also seems to be crucial in deriving well-performed strategies for voltage control and renewable accommodation.
\end{itemize}

\begin{figure}[!t]
    \centering
    \includegraphics[width=0.9\linewidth]{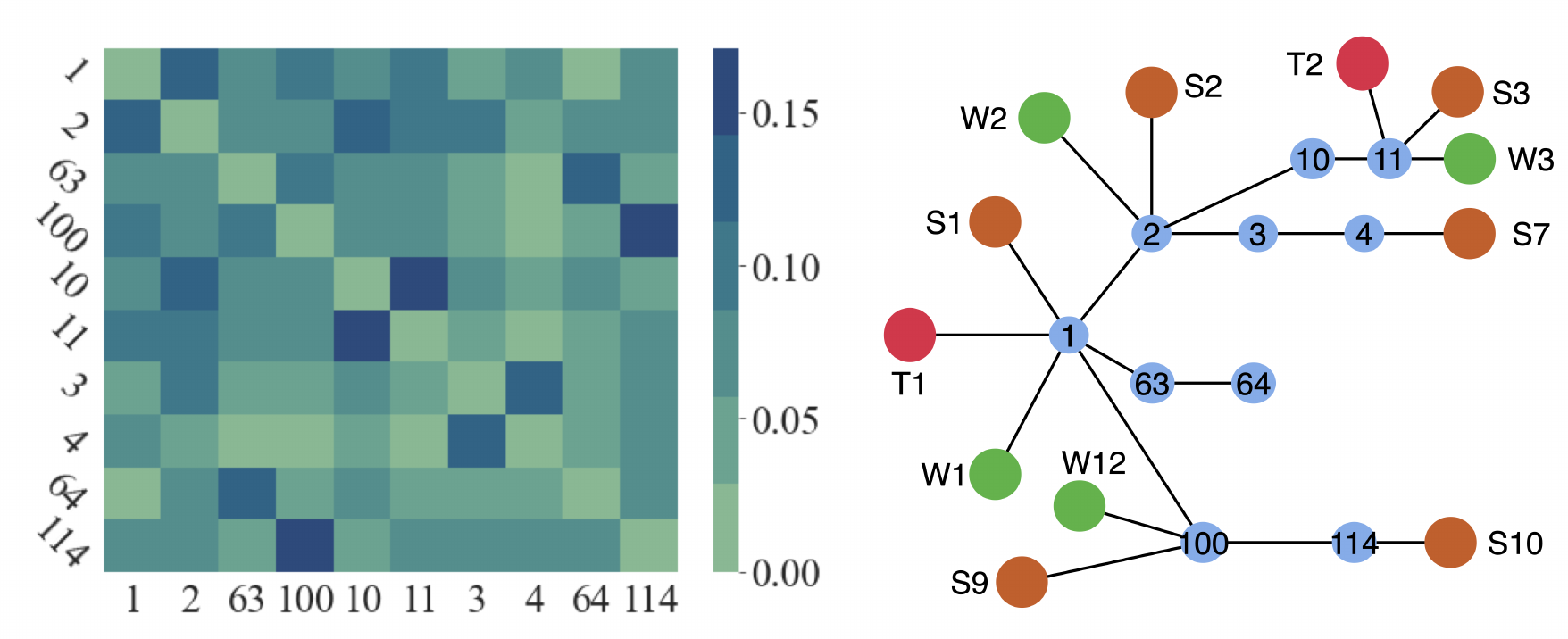}
    \caption{The partial spatial attention matrix and its corresponding sub-graph in IEEE 118-bus RDS.}
    \label{fig:satt_mat}
\end{figure}

\begin{figure}[!t]
    \centering
    \includegraphics[width=0.9\linewidth]{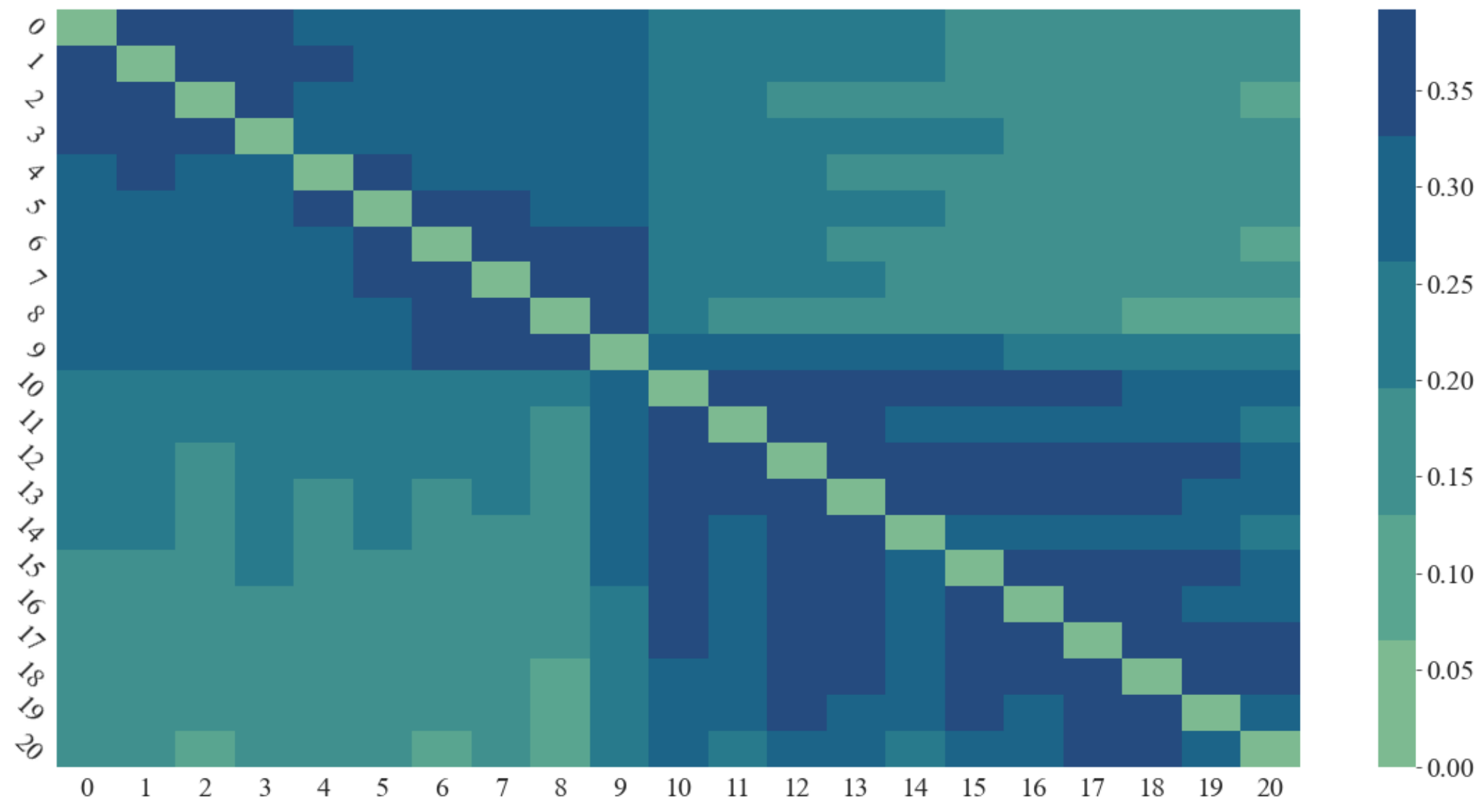}
    \caption{The partial temporal attention matrix of Bus $1$'s recent segment in IEEE 118-bus RDS.}
    \label{fig:tatt_mat}
\end{figure}

Additionally, we observe two interesting phenomena from spatial and temporal attention matrices, which are illustrated in Fig. \ref{fig:satt_mat} and Fig. \ref{fig:tatt_mat}, respectively.
\begin{itemize}
    \item In Fig. \ref{fig:satt_mat}, the spatial attention mechanism tends to focus on node pairs (larger attention weights) with more generator integration. For instance, although Bus $2$ and $114$ are not adjacent, they can be connected by Bus $1$ (connected with three generators) and $100$ (connected with two generators). Therefore, it is reasonable that the correlation strength between Bus $2$ and $114$ is more significant than other nonadjacent node pairs.
    \item In the temporal attention for the recent graph segment, the correlation strengths between current and historical node features drop sharply when it comes to the $10$-th historical time step, which may indicate that the latest $10$ feature vectors of one node contain more significant temporal correlation information.
\end{itemize}

\subsubsection{Stability Test of the DRL-based Strategy}  \label{subsubsec:exps_results_stability_test}
\begin{figure}[!t]
\centering
\subfloat[]{
\includegraphics[width=0.48\linewidth]{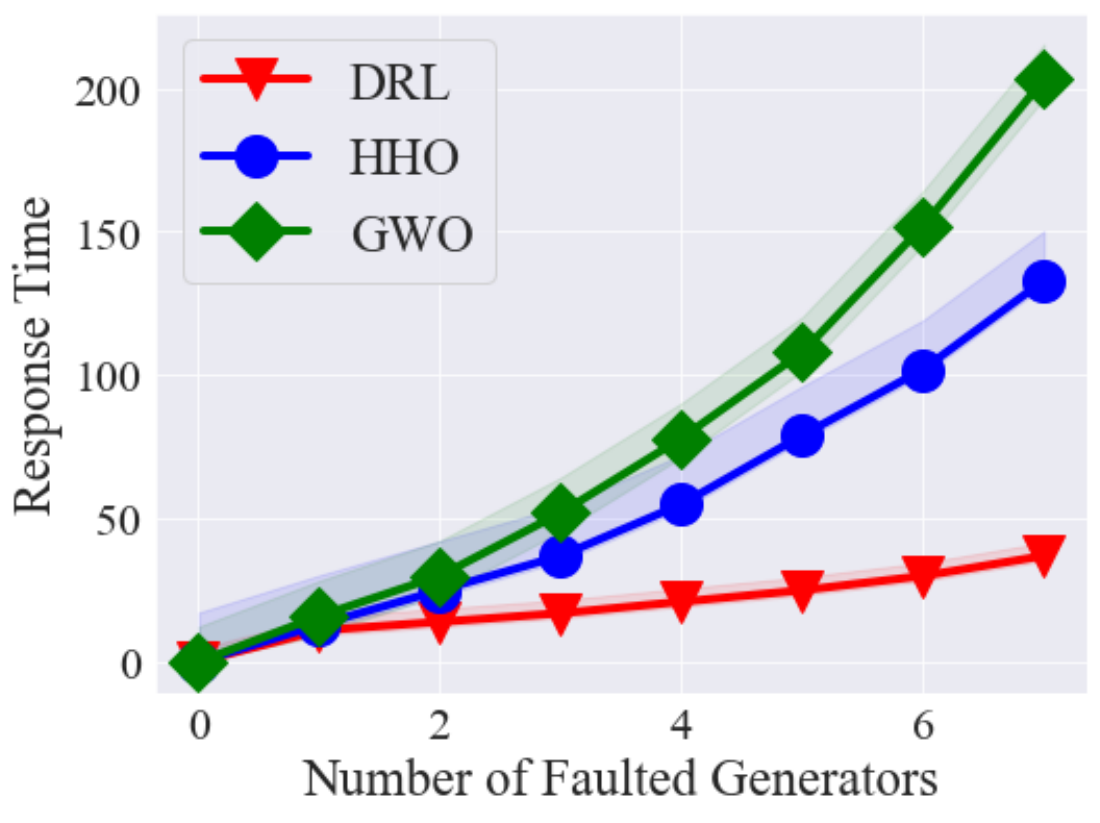}
\label{fig:different-response}
}
\subfloat[]{
\includegraphics[width=0.48\linewidth]{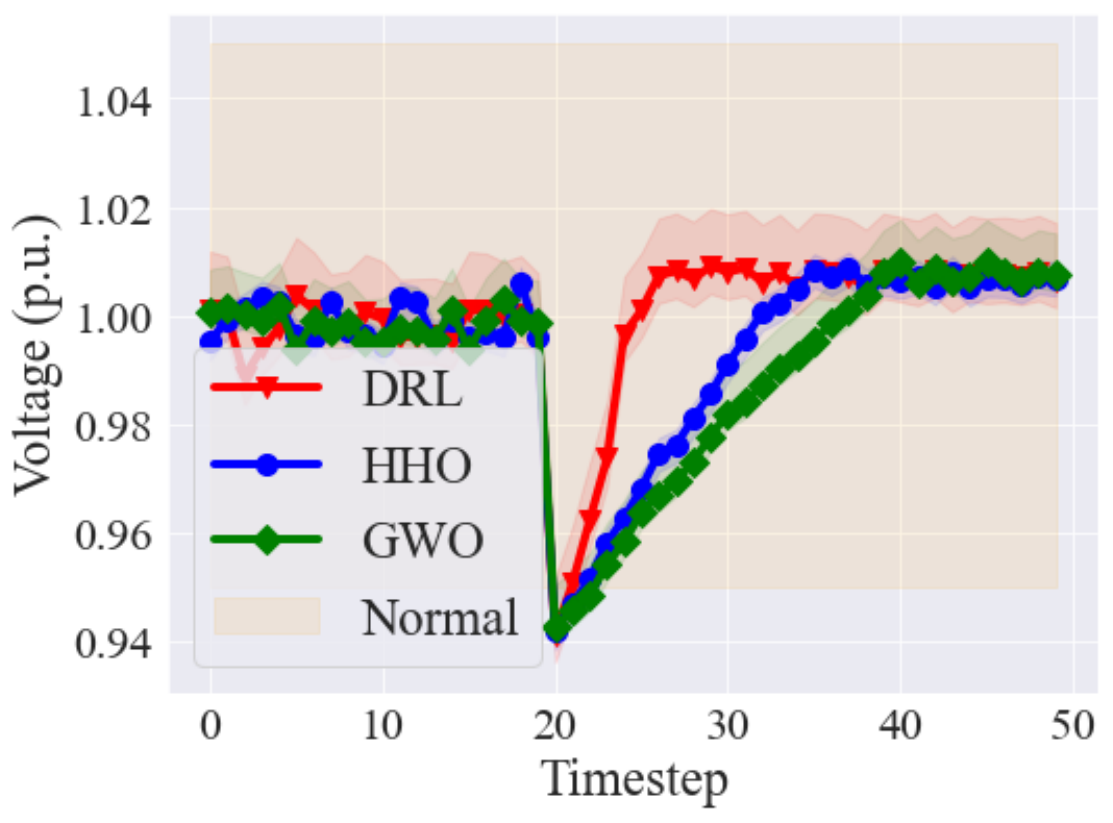}
\label{fig:specific-response}}
\caption{(a) Response time when facing different number of faulted generators in IEEE 69-bus RDS; (b) Case study of generator fault: one generator failed in IEEE 69-bus RDS.}
\label{fig:stability}
\end{figure}

We here define the average response time, calculated via counting how many time steps the DN takes to recover voltage back to its normal level, to assess the stability of the DN in the event of generator faults. Fig. \ref{fig:different-response} illustrates the response time of HHO, GWO, and the proposed DRL-based strategy with different numbers of faulted generators, where the DRL's response time grows linearly compared to the exponential growth of two heuristic algorithms. Besides, Fig. \ref{fig:specific-response} shows a more detailed case study with one faulted generator in the IEEE 69-bus RDS, indicating that the stability of DNs can be significantly improved through our DRL-based strategy.

\subsubsection{Trade-Off Between Voltage Fluctuation Control and Renewable Accommodation} \label{subsubsec:exps_results_voltage_control}
\begin{figure}[!t]
    \centering
    \includegraphics[width=0.7\linewidth]{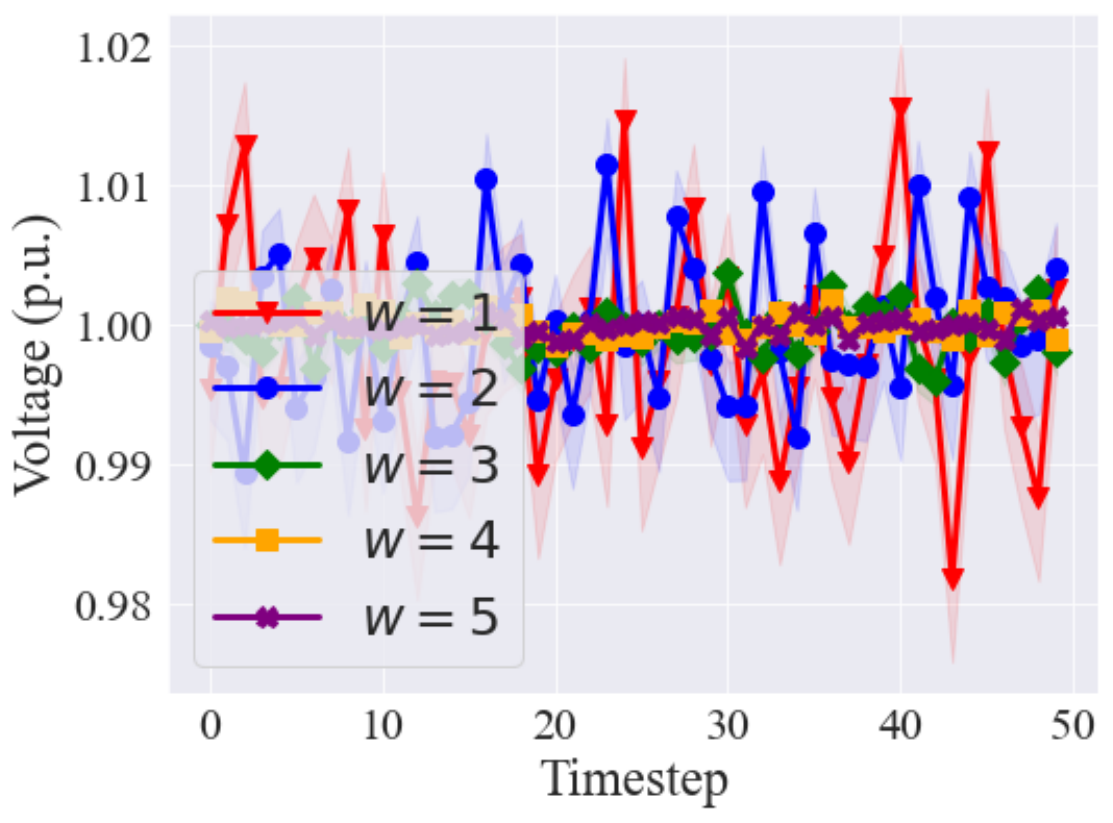}
    \caption{Voltage fluctuation of Bus $15$ in IEEE $69$-bus RDS.}
    \label{fig:voltage-fluctuation}
\end{figure}

The weights $w^\text{vol}$, $w^\text{RER}$, and $w^\text{gen}$ in our designed reward function represent their relative importance when learning the optimal control strategy. We trained and evaluated our DRL-based strategy with different voltage control weight $w^\text{vol}$ to investigate our method's capability in voltage fluctuation control. The results are presented in Table \ref{tab:weights}. Specifically, Bus $15$'s voltage fluctuation profiles in IEEE 69-bus RDS are illustrated in Fig. \ref{fig:voltage-fluctuation}. Surprisingly, better voltage control (with the increase of $w^\text{vol}$) leads to performance degradation of our DRL-based strategy, with lower SCOREs as shown in Table \ref{tab:weights}. Such results may suggest that overemphasizing voltage fluctuation control may undermine the strategy's performance. 

Furthermore, we found that, while the performance of voltage control $\alpha^\text{vol}$ improves with the increase of the weight coefficient $w^\text{vol}$, the renewable accommodation rate $\alpha^\text{RER}$ conversely decreases, as shown in Table \ref{tab:vol_weight}. Given that the DDPG's performance is related to both voltage control and renewable accommodation, the decreasing renewable accommodation rate may lead to the degradation of the DDPG's performance. Similarly, overemphasizing renewable accommodation also undermines DDPG's performance due to the increasing voltage fluctuation rates, as shown in Table \ref{tab:RER_weight}. Based on the evaluation results presented in Table \ref{tab:vol_weight} and \ref{tab:RER_weight}, we may conclude that controlling voltage fluctuation and accommodating renewable is a trade-off. Hence, striking an effective balance between voltage control and renewable accommodation seems to be the best solution. According to the parameter searching results presented in Table \ref{tab:weight_searching}, our current weight setting, i.e., $w^\text{vol}=1$, $w^\text{RER}=1$, and $w^\text{gen}=0.01$, seems to achieve the optimal balance between voltage control and renewable accommodation. Additionally, in real practice, if more strict voltage control is required, we can always increase the weight coefficient $w^\text{vol}$ to more effectively mitigate voltage fluctuations, though the renewable accommodation rate will inevitably decrease. Similarly, more renewable generation can be integrated into the DN with the increase of the weight coefficient $w^\text{RER}$.

\begin{table}[!t]
    \centering
    \caption{The SCOREs and voltage fluctuation rate with different voltage fluctuation control weight.}
    \begin{tabular}{c|| cc cc cc}
    $w^\text{vol}$  & \multicolumn{2}{c}{33-Bus} & \multicolumn{2}{c}{69-Bus} & \multicolumn{2}{c}{118-Bus}\\
    \hline
    Value & SCORE & $\alpha^\text{vol}$ & SCORE & $\alpha^\text{vol}$ & SCORE & $\alpha^\text{vol}$\\
    \hline
    $1$ & $4818$ & $0.22\%$ & $8533$ & $0.19\%$ & $15822$ & $0.20\%$\\
    \hline
    $2$ & $4775$ & $0.20\%$ & $8300$ & $0.18\%$ & $15352$ & $0.18\%$\\
    \hline
    $3$ & $4760$ & $0.18\%$ & $8394$ & $0.15\%$ & $15720$ & $0.16\%$\\
    \hline
    $4$ & $4814$ & $0.16\%$ & $8481$ & $\bm{0.14\%}$ & $15731$ & $0.15\%$\\
    \hline
    $5$ & $4824$ & $\bm{0.15\%}$ & $8513$ & $0.15\%$ & $15842$ & $\bm{0.14\%}$\\
    \hline
    \end{tabular}
    \label{tab:weights}
\end{table}

\begin{table}[!t]
    \centering
    \caption{The voltage fluctuation rates and renewable accommodation rates with different voltage fluctuation control weights.}
    \begin{tabular}{c|| cc cc cc}
    $w^\text{vol}$  & \multicolumn{2}{c}{33-Bus} & \multicolumn{2}{c}{69-Bus} & \multicolumn{2}{c}{118-Bus}\\
    \hline
    Value & $\alpha^\text{RER}$ & $\alpha^\text{vol}$ & $\alpha^\text{RER}$ & $\alpha^\text{vol}$ & $\alpha^\text{RER}$ & $\alpha^\text{vol}$\\
    \hline
    $1$ & $\bm{94.2\%}$ & $0.22\%$ & $\bm{93.4\%}$ & $0.19\%$ & $\bm{94.5\%}$ & $0.20\%$\\
    \hline
    $2$ & $90.8\%$ & $0.22\%$ & $91.4\%$ & $0.15\%$ & $91.0\%$ & $0.23\%$\\
    \hline
    $3$ & $88.2\%$ & $0.21\%$ & $89.9\%$ & $0.17\%$ & $89.5\%$ & $0.18\%$\\
    \hline
    $4$ & $86.6\%$ & $0.15\%$ & $86.1\%$ & $0.14\%$ & $87.2\%$ & $0.15\%$\\
    \hline
    $5$ & $84.3\%$ & $\bm{0.14\%}$ & $85.7\%$ & $\bm{0.15\%}$ & $86.1\%$ & $\bm{0.16\%}$\\
    \hline
    \end{tabular}
    \label{tab:vol_weight}
\end{table}

\begin{table}[!t]
    \centering
    \caption{The voltage fluctuation rates and renewable accommodation rates with different renewable accommodation weights.}
    \begin{tabular}{c|| cc cc cc}
    $w^\text{RER}$  & \multicolumn{2}{c}{33-Bus} & \multicolumn{2}{c}{69-Bus} & \multicolumn{2}{c}{118-Bus}\\
    \hline
    Value & $\alpha^\text{RER}$ & $\alpha^\text{vol}$ & $\alpha^\text{RER}$ & $\alpha^\text{vol}$ & $\alpha^\text{RER}$ & $\alpha^\text{vol}$\\
    \hline
    $1$ & $94.2\%$ & $\bm{0.22\%}$ & $93.4\%$ & $\bm{0.19\%}$ & $94.5\%$ & $\bm{0.20\%}$\\
    \hline
    $2$ & $93.7\%$ & $0.84\%$ & $93.0\%$ & $0.93\%$ & $94.2\%$ & $1.26\%$\\
    \hline
    $3$ & $93.7\%$ & $0.88\%$ & $93.5\%$ & $1.27\%$ & $95.1\%$ & $1.75\%$\\
    \hline
    $4$ & $94.0\%$ & $1.39\%$ & $93.6\%$ & $1.89\%$ & $95.6\%$ & $2.04\%$\\
    \hline
    $5$ & $\bm{94.5\%}$ & $1.52\%$ & $94.4\%$ & $1.92\%$ & $95.8\%$ & $2.31\%$\\
    \hline
    \end{tabular}
    \label{tab:RER_weight}
\end{table}

\section{Conclusion and Future Works}
\label{sec:conclusion}
In this paper, we proposed a DRL-based strategy to balance the trade-off between voltage fluctuations and renewable accommodation in the DN, with the aim of promoting the further adoption of RERs in the DNs. We first derive the optimization formulation considering voltage control and efficient renewable accommodation, along with generation cost minimization. A novel MG-ASTGCN is then proposed to fully explore the ST correlations among node pairs through ST attention mechanism and extract ST information of the DN through ST convolution. The extracted multi-time-scale ST information is finally delivered to the DDPG to learn the optimal control strategy. We can draw several conclusions based on experimental results: 1) extracting ST correlations in the DN plays an essential role in improving DDPG's performance and convergence speed, while the strategy's performance without ST attention mechanism or ST graphical information significantly degenerates; 2) compared with optimization-based benchmarks, the proposed DRL-based strategy achieves better performance along with less computation time. Besides, the adoption of the DRL-based strategy improves the network's stability in terms of shorter response time in the event of generator failures; 3) node pairs in the DN with more generator integration seem to have stronger spatial correlations. Moreover, in the temporal view, the most recent node feature vectors tend to contain more valuable temporal correlation information; 4) experimental results indicate that overemphasizing voltage fluctuation or renewable accommodation control may result in sub-optimal operations.

In our future work, designing a suitable incentive mechanism to accommodate more RERs in smart grids will be studied.


\bibliographystyle{IEEEtran}
\bibliography{IEEEabrv}

\end{document}